\documentclass[letterpaper,12pt]{article}
\pdfoutput=1

\usepackage{silence}
\usepackage{jheppub}
\usepackage{epsfig}
\usepackage{amsmath}
\usepackage{graphicx}
\usepackage{capt-of}
\usepackage{tikz-cd}
\usepackage{hyperref}

\usepackage[missing=]{gitinfo2}

\DeclareSymbolFont{AMSa}{U}{msa}{m}{n}
\DeclareSymbolFont{AMSb}{U}{msb}{m}{n}
\let\Box\relax
\DeclareMathSymbol{\Box}{\mathord}{AMSa}{"03}

\allowdisplaybreaks

\newcommand{\cX}{{\mathcal X}}
\newcommand{\cL}{{\mathcal L}}

\newcommand{\cM}{{\mathcal M}}

\newcommand{\cO}{{\mathcal O}}

\newcommand{\cT}{\mathcal T}

\newcommand{\C}{{\mathbb C}}

\newcommand{\bbZ}{{\mathbb Z}}

\newcommand{\bbR}{{\mathbb R}}
\newcommand{\Z}{{\bbZ}}
\newcommand{\R}{{\bbR}}

\newcommand{\tC}{\widetilde C}

\newcommand{\tW}{\widetilde W}

\newcommand{\lr}[2]{
	\langle #1,#2 \rangle
}

\newcommand{\insfigsvg}[3]{

\medskip
\noindent
\begin{minipage}{\linewidth}

\makebox[\linewidth]{\includegraphics[keepaspectratio=true,scale=#2]{figures/#1.pdf}}

\captionof{figure}{#3}

\label{fig:#1}
\end{minipage}
\medskip

}

\newcommand{\insfigpng}[3]{
	
	\medskip
	\noindent
	\begin{minipage}{\linewidth}
		
		\makebox[\linewidth]{\includegraphics[keepaspectratio=true,scale=#2]{figures/#1.png}}
		
		\captionof{figure}{#3}
		
		\label{fig:#1}
	\end{minipage}
	\medskip
	
}

\title{Exact WKB and the quantum Seiberg-Witten curve for 4d \texorpdfstring{$N=2$}{N=2} pure \texorpdfstring{$SU(3)$}{SU(3)} Yang-Mills,\\
Part I: Abelianization}

\author[1]{Fei Yan}
\affiliation[1]{NHETC and Department of Physics and Astronomy, Rutgers University}
\emailAdd{fyan.hepth@gmail.com}

\abstract{We investigate the exact WKB method for the quantum Seiberg-Witten curve of 4d $N=2$ pure $SU(3)$ Yang-Mills, in the language of abelianization. The relevant differential equation is a third-order equation on $\mathbb{CP}^1$ with two irregular singularities. We employ the exact WKB method to study solutions to such a third-order equation and the associated Stokes phenomena. We also investigate the exact quantization condition for a certain spectral problem. Moreover, exact WKB analysis leads us to consider new Darboux coordinates on a moduli space of flat SL(3,$\mathbb{C}$)-connections. In particular, in the weak coupling region we encounter coordinates of higher length-twist type generalizing Fenchel-Nielsen coordinates. The Darboux coordinates are conjectured to admit asymptotic expansions given by the formal quantum periods series; we perform numerical analysis supporting this conjecture.}

\begin{document}


\noindent{{\tiny \color{gray} \tt \gitAuthorIsoDate \gitAbbrevHash}}

\maketitle
\flushbottom

\section{Introduction}

Recently there has been much interesting progress in application of resurgence theory and exact WKB analysis to quantum mechanics (e.g.\cite{Alvarez_2000,Alvarez_2000_2,ZinnJustin:2004cg,ZinnJustin:2004ib,Jentschura_2010,Dunne:2013ada,Basar:2013eka,Dunne:2014bca,Misumi:2015dua,Behtash:2015loa,Dunne:2016qix,Fujimori:2017oab,Sueishi:2019xcj,Sueishi:2020rug,Sueishi:2021xti}). At the same time many interesting relations have been discovered between quantum mechanical systems and supersymmetric gauge theories (e.g.\cite{Nekrasov:2009rc,Alday:2009aq,Nekrasov:2011bc,Gaiotto:2014bza,Grassi:2014zfa,Marino:2015nla,Nekrasov:2015wsu,Basar:2015xna,Kashani-Poor:2015pca,Ashok:2016yxz,Basar:2017hpr,Codesido:2017jwp,Ito:2017iba,Ito:2018eon,Coman:2018uwk,Hollands:2019wbr,Grassi:2019coc,Ito:2020htm, Jeong:2020uxz,Haouzi:2020yxy,Lee:2020hfu,Coman:2020qgf,Imaizumi:2021cxf,Jeong:2021bbh}). These relations benefit both the study of supersymmetric gauge theories and quantum mechanical systems. In particular, they provide various tools to study the quantum mechanical problems in the context of quantization of Seiberg-Witten curves \cite{Seiberg:19941,Seiberg:19942} in certain 4d $N=2$ theories. Such tools are also very useful purely from the purpose to study differential equations, which could have broad applications, as an example see the recent developments in applying Seiberg-Witten techniques to black hole perturbation theory \cite{Aminov:2020yma,Bonelli:2021uvf,Bianchi:2021xpr,Bianchi:2021mft}. 

Our goal in this paper and its follow up \cite{Yan2}, is to study the quantum Seiberg-Witten curve of 4d $N=2$ pure $SU(3)$ super Yang-Mills (SYM) theory via different approaches. We hope our results not only benefit the study of super Yang-Mills theories, but also shed lights on the exact WKB analysis for higher order Schr\"odinger-like equations.

\subsection{Supersymmetric gauge theories and quantum mechanical systems}\label{sec:SUSYQM}

We begin with a review of correspondences between supersymmetric gauge theories and quantum mechanical systems from three different perspectives. 

\subsubsection{The  gauge/Bethe correspondence}
The first perspective arises in the context of gauge/Bethe correspondence \cite{Nekrasov:2009rc,Nekrasov:2009uh,Nekrasov:2009ui}, where 4d $N=2$ gauge theories in the $\Omega$-background provide the quantization of certain classical integrable systems. Concretely, one takes the Nekrasov-Shatashvili (NS) limit with $\epsilon_1=\hbar$ and $\epsilon_2\to 0$, the low energy effective theory is a 2d $N=(2,2)$ theory with an effective twisted superpotential:
\begin{equation}
\tW_{\text{eff}}(\textbf{a},\hbar)=\underset{\epsilon_2\to 0}{\text{lim}}\epsilon_2\text{log}Z(\textbf{a},\epsilon_1=\hbar,\epsilon_2),
\end{equation}
where $Z(\textbf{a},\epsilon_1,\epsilon_2)$ is the 4d Nekrasov partition function in $\Omega$-background \cite{Nekrasov:2002qd,Nekrasov:2003rj}. Here $\textbf{a}$ correspond to the vacuum expectation values of complex scalars in the $N=2$ vector multiplets. The effective 2d theory has a discrete set of vacua corresponding to solutions to 
\begin{equation}\label{eqn:2dvacua}
\text{exp}\left(\frac{\partial \tW_{\text{eff}}(\textbf{a},\hbar) }{\partial a_i}\right)=1, \quad i=1,\dots,r
\end{equation}
where $r$ is the rank of the 4d theory. In the context of quantum integrable systems, $\tW_{\text{eff}}(\textbf{a},\hbar)$ is identified with the Yang-Yang functional, and equation (\ref{eqn:2dvacua}) is the Bethe equation determining the set of eigenvalues of mutually commuting Hamiltonians. 
	
There is a special class of 4d $N=2$ theories called class $S $ theories  $\cT(\mathfrak{g},C)$, which are obtained by compactifying 6d $(2,0)$ theory of type $\mathfrak{g}$ on a Riemann surface $C$ with a partial topological twist \cite{Gaiotto:2009hg,Gaiotto:2009we}. Subjecting a class $S$ theory to $\Omega$-deformation in the NS limit \cite{Nekrasov:2009rc,Nekrasov:2010ka} gives quantization of the corresponding Hitchin integrable system where the phase space is the moduli space $\cM_H(\mathfrak{g},C)$\footnote{Here one needs to specify the global form of the gauge group $G$ corresponding to the Lie algebra $\mathfrak{g}$. In this notes we consider $\mathfrak{g}=sl(N)$ and $G=SU(N)$.} of solutions to Hitchin's equations on the Riemann surface $C$ \cite{hitchin1987}.

In this context, the Seiberg-Witten curve of the class $S$ theory is quantized to an {\it oper} \cite{2005math......1398B}, which is certain meromorphic differential operator on $C$. For example, take $\mathfrak{g}=A_{N-1}$, an oper in a class $S$ theory of type $A_{N-1}$ could be locally written as the following $N$-th order meromorphic operator on $C$:
\begin{equation}\label{eqn:SLNoper}
\cO_N(z):=\partial_z^N+t_2(z,\hbar)\partial_z^{N-2}+\dots+t_N(z,\hbar).
\end{equation}
Moreover, the variety of opers is a $\hbar$-dependent Lagrangian submanifold in $\cM_H(
\mathfrak{g},C,\hbar)$; it provides the quantization of Coulomb branch of the class $S$ theory.

In \cite{Nekrasov:2011bc} it was proposed that, for a class $S$ theory $\cT(A_{N-1},C)$, there exists a specific Darboux coordinate system, where the generating function for the variety of opers is identified with the effective twisted superpotential, up to certain boundary contribution at infinity. This is often called the NRS conjecture. As studied in \cite{Nekrasov:2011bc}, for $N=2$ such coordinates are complexified Fenchel-Nielsen coordinates \cite{DiscontinuousGroupsofIsometriesintheHyperbolicPlane}. Some examples of $N>2$ have been explored in \cite{Hollands:2017ahy,Jeong:2018qpc}, where the relevant Darboux coordinates are higher-rank analoges of Fenchel-Nielsen coordinates. In particular, \cite{Jeong:2018qpc} gave a gauge-theoretic derivation of NRS conjecture and its generalization to certain $A_2$ class-$S$ theory, by studying certain $1/2$-BPS codimension-two surface defects in the 4d theory. Concretely, the non-perturbative Dyson-Schwinger equation \cite{Nekrasov:2015wsu} satisfied by the surface defect partition function in $\Omega$-background gives a quantized version of opers; it reduces to the oper equation in the NS limit. 

\subsubsection{The topological string/spectral theory correspondence}

Another interesting development is the topological string/spectral theory (TS/ST) correspondence \cite{Grassi:2014zfa,Codesido:2015dia,Marino:2015nla}, which associates a non-perturbative quantum mechanical operator to a toric Calabi-Yau manifold. The relevant spectral problem is associated with the quantization scheme for mirror curves to the toric Calabi-Yau manifold. In particular, explicit expressions for the spectral determinant could be written down using a 11d version of the topological string free energy \cite{Grassi:2014zfa}, which allows the derivation of exact quantization condition for the operator spectrum. 

To make contact with 4d $N=2$ theories, one uses geometric engineering methods \cite{Katz:1996fh}, taking certain limit of topological string theory on appropriate Calabi-Yau geometry \cite{Grassi:2018bci,Grassi:2019coc}. Quantization of the Seiberg-Witten curve arises as the {\it canonical quantization} of an algebraic curve. 
As an example, \cite{Grassi:2018bci} studied a deformed Hamiltonian in quantum mechanics, which arises in the context of quantization of Seiberg-Witten curves (as hyperelliptic curves \cite{Argyres:1994xh,Klemm:1994qs,Klemm:1995wp,Seiberg:19941} ) for 4d $N=2$ pure $SU(N)$ SYM. A conjectural exact quantization condition was written down in closed form, using the 4d limit of TS/ST correspondence.

\subsubsection{Opers and the conformal limit}\label{sec:conformallimit}

There is yet another way, introduced in \cite{Gaiotto:2014bza}, to describe the quantization of the Coulomb branch of class $S$ theories in terms of the variety of opers. 

As described in \cite{Gaiotto:2009hg}, compactifying a class $S$ theory $\cT(\mathfrak{g},C)$ on a circle with radius $R$, the low energy effective theory is a 3d $N=4$ sigma model with target space being Hitchin moduli space $\cM_H(\mathfrak{g},C)$. Note that the radius $R$ of the compactification circle is a parameter in the corresponding Hitchin's equations. $\cM_H(\mathfrak{g},C)$ has a $\mathbb{CP}^1$-worth of complex structures parametrized by $\zeta$, where at $\zeta=0$ the moduli space is a complex integrable system. It is described as a torus fibration over the Coulomb branch of the 4d theory, where the torus fibre parameterize choice of Wilson lines and 't Hooft lines around the circle. With respect to the complex structure at $\zeta=0$, there is a canonical Lagrangian submanifold $\cL$, corresponding to the locus in $\cM_H(\mathfrak{g},C)$ where the Wilson and 't Hooft lines are turned off; $\cL$ is canonically isomorphic to the 4d Coulomb branch.

The quantization of 4d Coulomb branch in this setup happens in a special scaling limit, the so-called {\it conformal limit} \cite{Gaiotto:2014bza}, where one sends both $\zeta$ and $R$ to zero while keeping $\hbar=\zeta/R$ fixed. Viewing $\cM_H(\mathfrak{g},C)$ as a complex symplectic manifold in complex structure $\zeta$, the conformal limit is a well-defined scaling limit, and one denotes the resulting complex symplectic manifold as $\cM_H(\mathfrak{g},C;\hbar)$. Let $\cL_\hbar$ be the image of $\cL$ under the conformal limit, the statement is that $\cL_\hbar$ is a Lagrangian submanifold in $\cM_H(\mathfrak{g},C;\hbar)$, moreover it was conjectured to coincide with the variety of opers. The physical motivation behind this conjecture was through twisted compactification of the 4d theory on $\Omega_\hbar$-deformed cigar-like geometry \cite{Nekrasov:2010ka}, where $\cL_\hbar$ is related to the boundary condition at the tip of the cigar. This conjecture was demonstrated to be true in various examples, in \cite{Gaiotto:2014bza} and subsequent work such as \cite{Ito:2017ypt,Ito:2018eon,Hollands:2019wbr,Grassi:2019coc,Ito:2019llq,Dumas:2020zoz,Imaizumi:2021cxf}. From a mathematical point of view,  
this conjecture was proven in some special cases in \cite{2016arXiv160702172D}.

From a computational point of view, the conjecture in \cite{Gaiotto:2014bza} offers new insights to study the quantum Seiberg Witten curves of class $S$ theories. In particular, the conformal limit connects the exact WKB methods for opers with the exact WKB methods for flat connections parametrized by $R$ and $\zeta$; these methods were developed in \cite{Gaiotto:2009hg,Gaiotto:2012rg}. Concretely the Stokes graphs appearing in exact WKB for opers are the same as the {\it spectral networks} introduced in \cite{Gaiotto:2012rg}. As a result, techniques from the study of spectral networks could be directly applied to exact WKB analysis for Schr\"{o}dinger equations and their higher order analogues. This philosophy was further explored in \cite{Hollands:2019wbr}, which described the exact WKB method in the language of abelianization; we will give an overview of the method in Section \autoref{sec:WKBabelianization}.

\subsection{Different methods to compute resummed quantum periods}\label{sec:QPmethods}

In Section \autoref{sec:SUSYQM} we reviewed three setups where quantization of Seiberg-Witten curves occurs; this motivates different methods to compute the corresponding properly resummed {\it quantum periods}. Computing quantum periods using these different methods and checking their agreements play an important role in understanding the correspondences between 4d $N=2$ gauge theories and quantum mechanical systems. Recently such checks have been performed in 4d $N=2$ $SU(2)$ SYM without or with matter content, see e.g. \cite{Mironov:2009uv,He:2010if,Huang:2012kn,Basar:2015xna,Dunne:2016qix,Kashani-Poor:2015pca,Ashok:2016yxz,Ito:2017iba,Grassi:2019coc,Imaizumi:2021cxf,Grassi:2021wpw}. In this paper and \cite{Yan2}, we aim to generalize such analysis to higher rank theories, using 4d $N=2$ pure $SU(3)$ SYM as a concrete example.

Let us consider the quantum Seiberg-Witten curve for a class $S$ theory $\cT(A_{N-1},C)$, namely we look at the differential equation corresponding to an SL(N)-oper in (\ref{eqn:SLNoper}):
\begin{equation}\label{eqn:SLNoperODE}
\left[\partial_z^N+t_2(z,\hbar)\partial_z^{N-2}+\dots+t_N(z,\hbar)\right]\psi(z)=0.
\end{equation}
The standard WKB method makes the following ansatz for the wavefunction $\psi(z)$:
\begin{equation}\label{eqn:WKBansatzintro}
\psi(z)=\text{exp}\left(\frac{1}{\hbar}\int_{z_0}^z\lambda(z)dz\right)
\end{equation}
Substituting (\ref{eqn:WKBansatzintro}) into (\ref{eqn:SLNoperODE}) yields an order-N analogue of the Ricatti equation. The first step in constructing a solution is building a formal series solution to the Ricatti equation in powers of $\hbar$. At order-$\hbar^0$, the relevant equation describes the classical Seiberg-Witten curve for $\cT(A_{N-1},C)$, as a $N$-fold branched covering $\tC\to C$. There are $N$ choices of order-$\hbar^0$ solutions; they correspond to the $N$ sheets of the Seiberg-Witten curve $\tC$. We choose a sheet $i$ and consider the formal series solution
\begin{equation}
\lambda_i^{\text{formal}}(\hbar)=\sum\limits_{n=0}^\infty \lambda_i^{(n)}\hbar^n,
\end{equation}
where $\lambda_i^{(0)}$ is an order-$\hbar^0$ solution.
The higher order $\lambda_i^{(n)}$ are then uniquely fixed by recursively solving the Ricatti equation in orders of $\hbar$.

Classical periods of the Seiberg-Witten curve are given by the integrals of $\lambda^{(0)}$ along 1-cycles $\gamma$ of $\tC$, where $\gamma$ labels the IR electromagnetic (and flavor) charge. Correspondingly, the quantum periods or WKB periods are defined as
\begin{equation}\label{eqn:QPdef}
\Pi_\gamma(\hbar):=\oint_\gamma \lambda^{\text{formal}}(\hbar) dz, \quad \gamma\in H_1(\tC,\Z),
\end{equation}
where $\Pi_\gamma(\hbar)$ is a formal power series in $\hbar$
\begin{equation}\label{eqn:QPexpansion}
\Pi_\gamma(\hbar)=\sum\limits_{n=0}^\infty \Pi_\gamma^{(n)}\hbar^n, \quad \Pi_\gamma^{(n)}=\oint_\gamma \lambda^{(n)}dz.
\end{equation}

$\Pi_\gamma^{(n)}$ in general diverges as $n!$ \cite{VorosQuartic,Huang:2012kn,Basar:2015xna,Codesido:2016dld,Grassi:2019coc}; a natural way to properly resum $\Pi_\gamma(\hbar)$ is the Borel resummation. There could be rays in the Borel plane along which the Borel transform has singularities. As a consequence, $\Pi_\gamma(\hbar)$ is not Borel summable for certain phases of $\hbar$. One could nevertheless define lateral Borel resummations by slightly deforming the integration contour below or above the ray corresponding to such a phase. These two choices of deformation produce different answers, where the difference is defined as the {\it Stokes discontinuity} of quantum periods. Borel resummation of quantum periods and their associated Stokes discontinuities have a very rich structure, which has been an important topic in resurgence theory, see e.g. \cite{Aniceto:2018bis}. 

The correspondences between gauge theories and quantum mechanical systems reviewed in Section \autoref{sec:SUSYQM} motivate alternative ways to compute resummed quantum periods, which we describe below.

\subsubsection{Computing quantum periods via abelianization}\label{sec:abelianizationmethod}

Motivated by developments reviewed in \autoref{sec:conformallimit}, exact WKB methods could be reformulated geometrically in the context of {\it abelianization} \cite{Gaiotto:2012rg,Hollands:2013qza,Hollands:2017ahy,Hollands:2019wbr}, which maps a flat SL(N,$\C$)-connection over the Riemann surface $C$ to a flat GL(1,$\C$)-connection over the Seiberg-Witten curve $\tC\to C$. The Borel resummed quantum periods are closely related to the {\it Voros symbols} $\cX_\gamma(\hbar)$\footnote{More precisely the Voros symbol depends on a phase parameter $\theta$. Detailed definitions are described in Section \autoref{sec:WKBabelianization}.}, defined as holonomy of the flat abelian connection along 1-cycles $\gamma$ in $\tC$.

Using abelianization methods, $\cX_\gamma(\hbar)$ can be explicitly written as products of Wronskians of distinguished local solutions to the oper equation (\ref{eqn:SLNoperODE}). In this way $\cX_\gamma(\hbar)$ could be identified as {\it spectral coordinates} on a moduli space of flat SL(N,$\C$)-connections. In the case of SL(2)-opers, in generic situations $\cX_\gamma(\hbar)$ are Fock-Goncharov coordinates \cite{2003math.....11149F,Gaiotto:2009hg}; less generically one could also obtain exponentiated complexified Fenchel-Nielsen coordinates \cite{DiscontinuousGroupsofIsometriesintheHyperbolicPlane,Nekrasov:2011bc,Hollands:2013qza}. In the case of higher-rank SL(N)-opers, in special cases $\cX_\gamma(\hbar)$ could be identified with higher-rank Fock-Goncharov coordinates \cite{2003math.....11149F,Gaiotto:2012rg} or higher length-twist coordinates \cite{Hollands:2017ahy,Jeong:2018qpc} generalizing Fenchel-Nielsen coordinates. In general though, abelianization for higher-rank opers spells out  spectral coordinates that haven't been studied before.

The connection between $\cX_\gamma(\hbar)$ and quantum periods is mediated by certain asymptotic properties of $\cX_\gamma(\hbar)$ motivated from \cite{Gaiotto:2009hg,Gaiotto:2012rg,Gaiotto:2014bza,Hollands:2019wbr}. In particular, as $\hbar\to 0$ while staying within certain half of the $\hbar$-plane, $\text{log}\left(\cX_\gamma(\hbar)\right)$ admits an asymptotic expansion 
\begin{equation}\label{eqn:asymptotics}
\text{log}\left(\cX_\gamma(\hbar)\right)\sim \frac{1}{\hbar}\Pi_\gamma(\hbar),
\end{equation}
where $\Pi_\gamma(\hbar)$ is the formal series of quantum periods defined in (\ref{eqn:QPdef}). Moreover if $\hbar\to 0$ along the central ray within the half-plane, $\hbar\text{log}\left(\cX_\gamma(\hbar)\right)$ produces the Borel resummed quantum periods.\footnote{If $\Pi_\gamma(\hbar)$ happens to be not Borel-summable, then $\hbar\text{log}\left(\cX_\gamma(\hbar)\right)$ is conjectured to produce the median Borel summation of $\Pi_\gamma(\hbar)$ \cite{Hollands:2019wbr}.} 

For SL(2)-opers, mathematically speaking such asymptotic behavior of Voros symbols has been proven by Koike-Sch\"{a}fke and further studied by \cite{nikolaev2019abelianisation,Allegretti:2018kvc,Allegretti:2020dyt}; for higher SL(N)-opers though it remains to be a conjecture. Certain numerical evidence for this conjecture in the higher rank case had been provided in \cite{Hollands:2019wbr,Dumas:2020zoz}. In this paper we provide further evidence in the concrete example of an SL(3)-oper which appears in the quantization of Seiberg-Witten curve of 4d $N=2$ pure $SU(3)$ SYM.

\subsubsection{Computing quantum periods via TBA-like integral equations}\label{sec:TBAmethod}

The spectral coordinates $\cX_\gamma(\hbar)$ obey certain TBA-like integral equations \cite{Gaiotto:2014bza}, which could be viewed as the conformal limit of the TBA-like integral equations in \cite{Gaiotto:2008cd}. Such integral equations are determined by the BPS spectrum of the corresponding 4d $N=2$ theory; they are very useful in the study of quantum periods. 

There exists an interesting correspondence between BPS states and resurgent properties of quantum periods \cite{Gaiotto:2014bza,Ito:2018eon,Grassi:2019coc,Ito:2019llq}. Singularities of Borel transform for quantum periods are controlled by BPS spectrum of the 4d theory. Moreover the Stokes discontinuities of quantum periods are closely related to the Kontsevich-Soibelman transformation \cite{Kontsevich:2008fj,Gaiotto:2008cd,Gaiotto:2009hg}. As a consequence, BPS spectrum and subsequently the integral equations predict the locations of singularities in the Borel plane and the discontinuites in lateral Borel resummations of quantum periods.

The Borel resummed quantum periods are solutions to the integral equations; in principal one can compute them by solving the integral equations iteratively, after specifying appropriate boundary conditions on the solutions. This perspective has been explored in various examples in \cite{Gaiotto:2014bza,Grassi:2014zfa,Ito:2018eon,Hollands:2019wbr,Grassi:2019coc,Dumas:2020zoz}. In upcoming \cite{Yan2}, we provide numerical calculation of resummed quantum periods using the integral equations.

\subsubsection{Computing quantum periods via instanton calculus}\label{sec:Instantonmethod}

The NS limit \cite{Nekrasov:2009rc} of instanton calculus \cite{Nekrasov:2002qd,Nekrasov:2003rj} also provides a resummation of quantum periods \cite{Mironov:2009uv,Grassi:2019coc}. Concretely instanton calculous picks up distinguished quantum periods: the quantum $A$- and $B$-periods $\{a_1(\hbar),\dots,a_r(\hbar),a_D^1(\hbar),\dots, a_D^r(\hbar)\}$\footnote{Here $r$ denotes the rank of the theory.} satisfying the quantum special geometry relation
\begin{equation}\label{eqn:aDformula}
a_D^i(a_1,\dots,a_r;\hbar)=\frac{\partial F_{\text{NS}}(a_1,\dots,a_r;\hbar)}{\partial a_i}, \quad i=1,\dots,r
\end{equation}
Here $F_{\text{NS}}(a_1,\dots,a_r;\hbar)$ is the Nekrasov-Shatashvili free energy \cite{Nekrasov:2009rc}, which is $\hbar$ times the effective twisted superpotential $\tW_{\text{eff}}$. This free energy is given as a power series in the instanton counting parameter with a non-vanishing convergence radius in certain parameter range around the semiclassical region. 

Given the NS free energy, the quantum $A$-periods $a_i(\hbar)$ could be obtained by inverting the quantum Matone relation \cite{Matone:1995rx,Losev:2003py,Flume:2004rp,Aganagic:2011mi,Basar:2015xna,Fucito:2015ofa,Bullimore:2014awa,Basar:2017hpr,Codesido:2017jwp}. The quantum $B$-periods $a_D^i(\hbar)$ are then computed via (\ref{eqn:aDformula}). For appropriate parameter range, both $a_i(\hbar)$ and $a_D^i(\hbar)$ are convergent series expansion in the counting parameter with non-zero convergence radius, in particular they are exact in $\hbar$. Instanton calculus thus provides a natural resummation for the quantum $A$- and $B$-periods.\footnote{Relation between instanton resummation and Borel resummation has been clarified in \cite{Grassi:2019coc}, in the context of $N=2$ pure $SU(2)$ SYM.}  

The NRS conjecture and its higher rank generalization \cite{Nekrasov:2011bc,Hollands:2017ahy,Jeong:2018qpc} suggests that the instanton resummed quantum $A$- and $B$-periods correspond to certain special spectral coordinates: the Fenchel-Nielsen coordinates or higher length-twist coordinates. Symbolically\footnote{There are certain ambiguities in FN or higher length-twist coordinates, corresponding to a certain monodromy action mixing $a$ with $a_D$. We thank Andrew Neitzke for discussions on this.} one expects the following relation in the higher rank case:
\begin{equation}
\text{log}\left(\cX_\gamma^{\text{length}}(\hbar)\right)=\frac{1}{\hbar}a(\hbar), \quad 
\text{log}\left(\cX_\gamma^{\text{twist}}(\hbar)\right)=\frac{1}{\hbar}a_D(\hbar).
\end{equation}

\subsection{The canonical quantization and surface defects}

Up until now we have avoided talking about an important issue in the quantization of Seiberg-Witten curves. In principal there could be different quantization choices which reduce to the same classical Seiberg-Witten curve. A natural question would be which choice corresponds to the {\it canonical} quantization, where observables such as the Borel resummed quantum periods agree with predictions from TBA equations and instanton calculus. There has been many discussions on the canonical quantum Seiberg-Witten curve from gauge theory considerations \cite{,Nekrasov:2009rc,Nekrasov:2010ka,Kozlowski:2010tv,Mironov:2009uv,Alday:2009aq,Gaiotto:2009ma,Marshakov:2009gn,Mironov:2009by,Mironov:2009dv,Wyllard:2009hg,Alday:2009fs,Maruyoshi:2010iu,Poghossian:2010pn,Popolitov:2010bz,Zenkevich:2011zx,Basar:2015xna,Kashani-Poor:2015pca,Ashok:2016yxz,Basar:2017hpr,Ito:2017iba,Grassi:2018bci,Jeong:2018qpc,Jeong:2021bbh}. In upcoming \cite{Yan2}, we derive such a canonical quantization using the canonical surface defect instanton partition function, along the lines of \cite{Jeong:2018qpc}. As a further remark, canonical surface defect also plays an important role in the exact WKB analysis for the quantum Seiberg-Witten curve. In particular, the soliton spectrum of the surface defect determines the Stokes curves governing the Stokes phenomena for solutions to the differential equation.

There is another point of view about the canonical quantization. The classical Seiberg-Witten curve arises here in the class-$S$ construction of $N=2$ pure $SU(3)$ super Yang-Mills (SYM) theory. Geometrically the canonical choice of quantum Seiberg-Witten curve should come from analyzing the conformal limit \cite{Gaiotto:2014bza} of the Hitchin section \cite{hitchin1987} in the corresponding Hitchin moduli space. Such limit for the case of $SL(N)$-opers on a Riemann surface without any punctures has been analyzed in \cite{2016arXiv160702172D}. The conformal limit of Hitchin section for the case of punctured Riemann surfaces is in general not well-understood, even more so in presence of irregular punctures\footnote{For regular punctures, the canonical choice is the one under which the monodromies around regular punctures are unipotent.}. Our strategy here will be a bit experimental; we will give two possible quantization choices in (\ref{eqn:SU3eqn}), where both choices are motivated from \cite{2016arXiv160702172D}. As we will see, asymptotic analysis for spectral coordinates confirms both choices in (\ref{eqn:SU3eqn}) are valid quantizations for the classical class-$S$ Seiberg-Witten curve of pure $SU(3)$ SYM. However only one choice is {\it canonical} from the gauge theory point of view, as we will derive in upcoming \cite{Yan2}.

\subsection{The \texorpdfstring{$SU(3)$}{SU3} equation}

In this paper and its followup \cite{Yan2}, we investigate the  quantum Seiberg-Witten curve for the $N=2$ pure SU(3) super Yang-Mills. Work exploring some perspectives of this theory has appeared in e.g. \cite{Popolitov:2010bz,Grassi:2018bci,Fioravanti:2019awr}. Here we start with the canonical Seiberg-Witten curve appearing in the class-$S$ construction of $N=2$ pure $SU(3)$ SYM, via compactifying a 6d $(2,0)$ theory of type $A_2$ on $\mathbb{CP}^1$ with two irregular singularities at $z=0$ and $z=\infty$.

We consider two possible quantizations of this Seiberg-Witten curve in terms of the following third-order differential equations:
\begin{equation}\label{eqn:SU3eqn}
\begin{split}
&A: ~ \left[\partial_z^3+\hbar^{-2}\frac{u_1}{z^2}\partial_z+\left(\hbar^{-3}\left(\frac{\Lambda}{z^4}+\frac{u_2}{z^3}+\frac{\Lambda}{z^2}\right)-\hbar^{-2}\frac{u_1}{z^3}\right)\right]\psi(z)=0,\\
&B: ~ \left[\partial_z^3+\hbar^{-2}\frac{u_1+\hbar^2}{z^2}\partial_z+\left(\hbar^{-3}\left(\frac{\Lambda}{z^4}+\frac{u_2}{z^3}+\frac{\Lambda}{z^2}\right)-\hbar^{-2}\frac{u_1+\hbar^2}{z^3}\right)\right]\psi(z)=0,
\end{split}
\end{equation}
where $u_1$ and $u_2$ are Coulomb branch parameters. We denote both equations as the $SU(3)$ equations; they both produce Voros symbols with the expected asymptotics. However equation $B$ is the canonical quantum Seiberg-Witten curve, as we describe more in \cite{Yan2}.

As discussed in \autoref{sec:QPmethods}, there are different methods to compute properly resummed quantum periods in this theory. In this paper we focus on the method described in \autoref{sec:abelianizationmethod}, namely we compute Borel resummed quantum periods via abelianization. In the followup \cite{Yan2}, we investigate the TBA method described in \autoref{sec:TBAmethod} and the instanton calculus method described in \autoref{sec:Instantonmethod}, along with perspectives from the instanton partition function of the $N=2$ $SU(3)$ Yang-Mills with the insertion of a canonical $1/2$-BPS codimension-two surface defect.

Concretely we study loci in both the strong-coupling region and the weak-coupling region. In the strong-coupling chamber with 12 BPS states, the Voros symbol $\cX_\gamma(\hbar)$ could be expressed in terms of Wronskians of distinguished local solutions to (\ref{eqn:SU3eqn}), which decay exponentially as one goes into the irregular singularity at $z=0$ or $z=\infty$. $\cX_\gamma(\hbar)$ are certain coordinates for flat SL(3,$\C$)-connections over $\mathbb{CP}^1$ with two irregular singularities; such coordinates haven't been considered before as far as we know. We also numerically evaluate $\text{log}\left(\cX_\gamma(\hbar)\right)$ and compare with the expected asymptotic quantum periods expansion to certain order in $\hbar$. We find good numerical agreement; we view this as evidence that higher-order exact WKB analysis via abelianization does work as expected.

We also investigate certain loci in the weak-coupling region, where $\cX_\gamma(\hbar)$ is expressed using exponentially decaying local solutions as one goes into an irregular singularity, as well as eigenvectors of the monodromy around the irregular singularity. The $\cX_\gamma(\hbar)$ constructed in such loci of the weak-coupling region is an instance of the higher length-twist coordinates. Different from examples in \cite{Hollands:2017ahy,Jeong:2018qpc}, here $\cX_\gamma(\hbar)$ are coordinates on a moduli space of flat SL(3,$\C$)-connections over a surface containing irregular singularities. We perform numerical check against the expected asymptotic quantum periods expansion. Additionally we comment on the exact quantization condition (EQC) for a bound state problem associated with the differential equation (\ref{eqn:SU3eqn}).

Finally we remark that, it would be very interesting to have a gauge-theoretical derivation of higher length-twist coordinates in this example and understand the generalized NRS conjecture, by studying the canonical surface defect in the $N=2$ pure $SU(3)$ theory, following the work of \cite{Jeong:2018qpc}.

\section{Exact WKB and abelianization for \texorpdfstring{SL(3)}{SL3}-opers}\label{sec:WKBabelianization}

The exact WKB method is an approach to study the Stokes data of linear scalar differential equations. Originating from the study of Schr\"{o}dinger equations, it has been developed in large amount of literature, for a sampling of literature see e.g. \cite{VorosQuartic,Voros1983,Silverstone,Delabaere1997ExactSE,KawaiTakeioriginal,Iwaki2014ExactWA}. Recently exact WKB method for higher order Schr\"{o}dinger-like equations has also been initiated in e.g. \cite{aoki1991new,Aoki_2005,2008ShudoIkeda,Honda:2040784,SASAKI2016711}.

In \cite{Gaiotto:2012rg,Hollands:2019wbr} a new geometric reformulation of exact WKB has been proposed, for Schr{\"o}dinger operators and higher order opers. A key ingredient in this reformulation is a process of {\it abelianization}, which maps a flat SL(N,$\C$)-connection over a Riemann surface $C$ to a flat GL(1,$\C$)-connection over a $N$-fold covering $\tC \to C$. In this section we review exact WKB analysis in the language of abelianization, following closely the descriptions in \cite{Hollands:2019wbr}. 

Concretely we focus on the case of SL(3)-opers, namely an third-order differential equation involving two meromorphic potentials $P_2(z,\hbar)$ and $P_3(z,\hbar)$:
\begin{equation}\label{eqn: order3oper}
\left[\partial_z^3+\hbar^{-2}P_2(z,\hbar)\partial_z+\left(\hbar^{-3}P_3(z,\hbar)+\frac{1}{2}\hbar^{-2}P_2'(z,\hbar)\right)\right]\psi(z)=0.
\end{equation}
Although this equation is written explicitly in a single coordinate patch, it could be formulated on a Riemann surface $C$ with a complex projective structure. In that context, $\psi(z)$ is interpreted as a section of $K_C^{-1}$ where $K_C$ is the canonical bundle, $P_2$ is a meromorphic quadratic differential while $P_3$ is a meromorphic cubic differential. In particular, equation (\ref{eqn: order3oper}) could be viewed as the quantization of the following Seiberg-Witten curve:
\begin{equation}
\tC=\{\lambda: \lambda^3+p_2\lambda+p_3=0\}\subset T^*C,
\end{equation}
where $\lambda$ denotes the Seiberg-Witten differential, $p_{2,3}$ are the order-$\hbar^{0}$ terms in $P_{2,3}$.

The exact WKB method for SL(3)-opers (or higher SL(N)-opers) is not yet on solid footing in mathematics, however it was conjectured in \cite{Hollands:2019wbr} that the traditional exact WKB method for Schr{\"o}dinger equations could be extended to SL(3)-opers, by combining the methods developed in \cite{Gaiotto:2012rg} with the scaling limit of \cite{Gaiotto:2014bza}. Numerical evidence supporting this conjectural picture in certain examples have appeared in \cite{Hollands:2017ahy,Hollands:2019wbr,Dumas:2020zoz}. Our description here also builds upon this conjecture; and we will provide further numerical evidence supporting this conjecture in Sections \autoref{cha: SU3strong} and \autoref{cha: SU3weak}.

\subsection{The WKB solutions}\label{sec:WKBsol}

The exact WKB method is centered around construction of distinguished local WKB solutions, see e.g. \cite{AIHPA_1983__39_3_211_0,Delabaere1997ExactSE,Iwaki2014ExactWA} for the case of Schr{\"o}dinger equations.

Written in local coordinate $z$ on a contractible open set $U\subset C$, a WKB solution of (\ref{eqn: order3oper}) on $U$ takes the following form:
\begin{equation}\label{eqn: WKBansatz}
\psi(z)=\text{exp}\left(\frac{1}{\hbar}\int_{z_0}^z\lambda(z)dz\right),
\end{equation} 
where $z_0\in U$ is a chosen basepoint. For $\psi(z)$ to be a solution of (\ref{eqn: order3oper}), $\lambda(z)$ must obey the following third-order analogue of the Riccati equation:
\begin{equation}\label{eqn: Riccati}
\lambda^3(z)+3\hbar\lambda(z)\partial_z\lambda(z)+\hbar^2\partial_z^2\lambda(z)+P_2(z,\hbar)\lambda(z)+P_3(z,\hbar)+\frac{1}{2}\hbar P_2'(z,\hbar)=0.
\end{equation}

We first build a formal series solution $\lambda^{\text{formal}}$ in powers of $\hbar$. At order-$\hbar^0$, the Riccati equation (\ref{eqn: Riccati}) becomes 
\begin{equation}\label{eqn: orderh0}
\left(\lambda^{(0)}\right)^3+p_2\lambda^{(0)}+p_3=0,
\end{equation}
where $\lambda^{(0)}$ is the leading order-$\hbar^0$ term in the formal series. Thus we encounter a 3-fold ambiguity; this could be resolved by choosing a solution to (\ref{eqn: orderh0}), or equivalently by choosing a sheet $i$ of the following 3-fold covering of $C$:
\begin{equation}\label{eqn: 3foldcover}
\tC=\{\lambda^{(0)}: \left(\lambda^{(0)}\right)^3+p_2\lambda^{(0)}+p_3=0\}.
\end{equation}
In WKB language this is usually called the {\it WKB curve}, which could be identified with the Seiberg-Witten curve of a 4d $\mathcal{N}=2$ theory; in particular $\lambda^{(0)}$ corresponds to the Seiberg-Witten differential.

Once we have chosen a sheet $i$ and the corresponding solution $\lambda^{(0)}_i$ to (\ref{eqn: orderh0}), higher order terms in the formal series are determined by solving (\ref{eqn: Riccati}) perturbatively in $\hbar$. For example, suppose we take $P_2=0$ and $P_3$ doesn't depend on $\hbar$, then the first few orders in this formal series take the form
\begin{equation}
\begin{split}
\lambda^{\text{formal}}_i&=\sum\limits_{n=0}^\infty\lambda_i^{(n)}\hbar^n\\
&=\lambda_i^{(0)}-\hbar \frac{P_3'}{3P_3}+\hbar^2\frac{6 P_3 P_3''-7\left(P_3'\right)^2}{27P_3^2 \lambda_i^{(0)}}+\dots
\end{split}
\end{equation}

By construction $\lambda_i^{\text{formal}}$ is only a formal series in $\hbar$, substituting $\lambda_i^{\text{formal}}$ into the WKB ansatz (\ref{eqn: WKBansatz}) we obtain formal solutions $\psi_i^{\text{formal}}$ to the differential equation (\ref{eqn: order3oper}). A meaningful question is: under what conditions could one interpret $\psi_i^{\text{formal}}$ as an asymptotic series of certain actual solution $\psi_i$, such that $\psi_i\sim\psi_i^{\text{formal}}$ as $\hbar\to 0$? A natural way to produce the solutions $\psi_i$ is to perform Borel resummation of $\psi_i^{\text{formal}}$. However in general one can not do so {\it globally} on $C$, but only away from the so-called {\it Stokes curves}. The Stokes curves are one-dimensional curves on the Riemann surface $C$; they divide $C$ into different regions. Within each region Borel resummation produces actual solutions $\psi_i$, which jump as one goes across a Stokes curve.

\insfigsvg{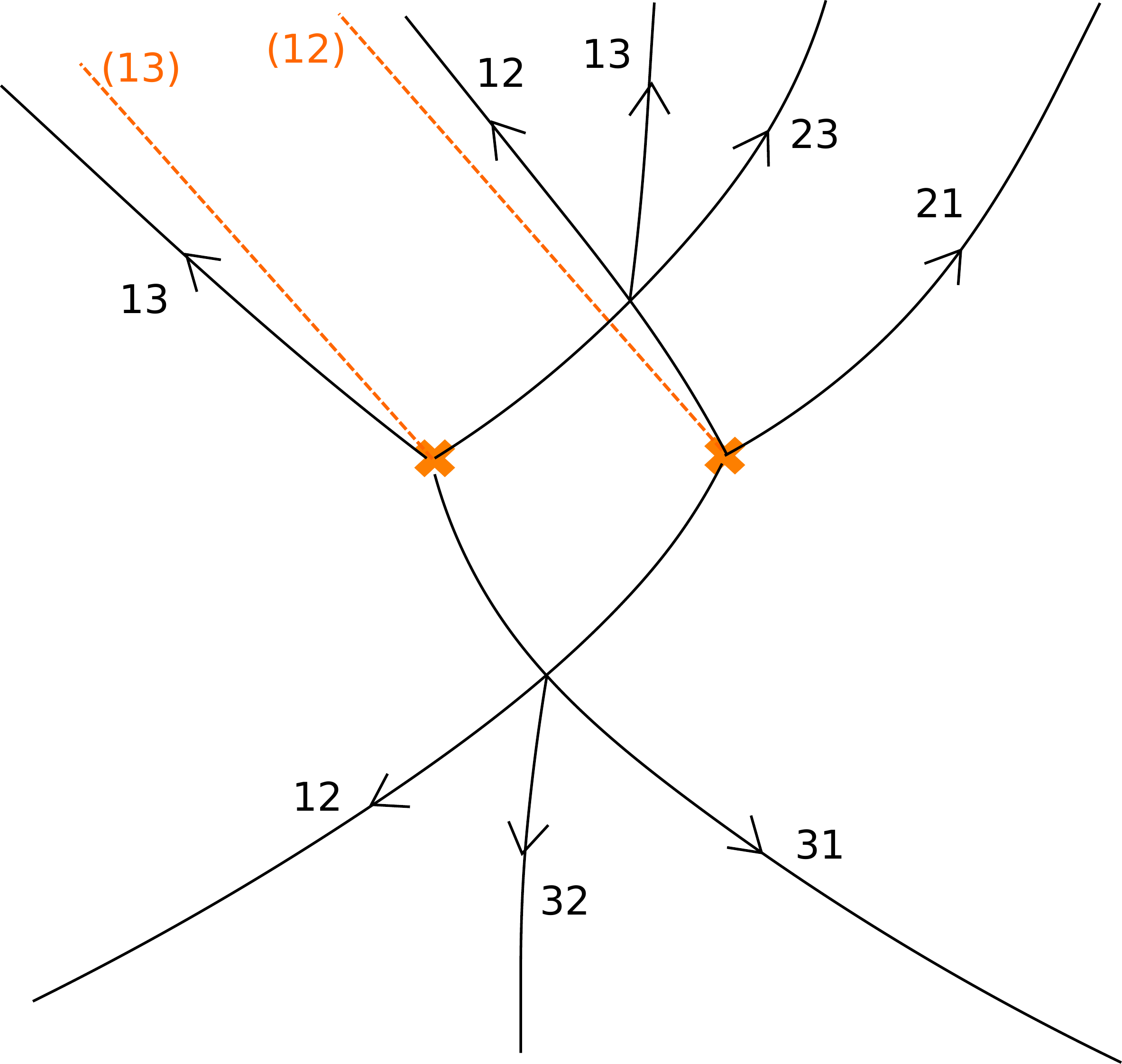}{0.22}{A representative local picture of a Stokes graph in the neighborhood of two simple branch points of different types. The orange crosses represent simple branch points, while the orange dashed lines represent the choice of branch cuts. }

In fact one can define Stokes curves more generally as {\it $\theta$-Stokes curves} labeled by a phase $\theta$. The conjectural picture is, away from such curves there exist actual solutions $\psi_i^\theta$, such that $\psi_i^\theta$ has the desired expansion $\psi_i^\theta\sim \psi_i^{\text{formal}}$ as $\hbar\to 0$ while staying within the following half-plane
\begin{equation}\label{eqn:Hthetaplane}
\mathbb{H}_\theta=\{\hbar: \text{Re}\left(\text{e}^{-\text{i}\theta}\hbar\right)> 0\}.
\end{equation}
Taking $\theta=\text{Arg}(\hbar)$, one then obtains the Stokes curves appearing in the Borel resummation story. 

The $\theta$-Stokes curves are one-dimensional curves on the Riemann surface $C$ carrying labels $ij$. They are defined as follows: along an oriented $\theta$-Stokes curve of type $ij$, $\text{e}^{-\text{i}\theta}\left(\lambda_i^{(0)}-\lambda_j^{(0)}\right)dz$ is real and positive, where $\lambda_{i(j)}^{(0)}$ are the $i(j)$-th solution to the WKB curve (\ref{eqn: orderh0}). We remark that the above condition could be interpreted as the BPS conditions for solitons associated with the canonical surface defect in the corresponding class-$S$ theory \cite{Gaiotto:2012rg}. In this sense, the BPS soliton spectrum controls the Stokes phenomena for solutions to the quantum Seiberg-Witten curve, such as equation (\ref{eqn: order3oper}).

For simplicity suppose the 3-fold covering $\tC$ only has simple branch points, then from each branch point there emanate three $\theta$-Stokes curves. An important new feature of $\theta$-Stokes curves for higher order SL(N)-opers, is that a $\theta$-Stokes curve of type $ik$ could be born from the intersection of $\theta$-Stokes curves of types $ij$ and $jk$ \cite{aoki1991new,Aoki_2005,Gaiotto:2012rg}. The collection of $\theta$-Stokes curves defines the {\it Stokes graph} $W$ at phase $\theta$. Examples of Stokes graphs for the $SU(3)$ equation could be found in \autoref{fig:smallu1}, \autoref{fig:strong-Stokes} and \autoref{fig:weak-Stokes}. In \autoref{fig:localStokes} we give a representative local picture of a Stokes graph in the neighborhood of two simple branch points.

Within a domain that doesn't contain any $\theta$-Stokes curves of type $ij$, one obtains actual solutions $\psi_i^\theta(z)$ and $\psi_j^\theta(z)$ to the SL(3)-oper equation (\ref{eqn: order3oper}). In particular if the domain doesn't contain any $\theta$-Stokes curves of any type, we would obtain three solutions $\psi^\theta_{1,2,3}(z)$, which give a basis of solutions to the SL(3)-oper equation (\ref{eqn: order3oper}). As one crosses a $\theta$-Stokes curve of type $ij$, the solution $\psi_i^\theta$ jumps by a constant multiple of $\psi_j^\theta$. The rules for gluing together solutions across $\theta$-Stokes curves could be formulated in the context of {\it abelianization} \cite{Gaiotto:2012rg,Hollands:2013qza}, which we describe below.

\subsection{Abelianization}\label{sec:gluingrules}

The SL(3)-oper equation (\ref{eqn: order3oper}) could be written as a first order differential equation with $3\times 3$ matrix coefficients, concretely in a local patch such equation is given as
\begin{equation}\label{eqn:SL3matrixoper}
\left[\partial_z+\hbar^{-1}\begin{pmatrix}
0 & -\frac{\sqrt{2}}{4}P_2 & \frac{P_3}{2}\\
\sqrt{2} & 0 & -\frac{\sqrt{2}}{4}P_2\\
0 & \sqrt{2} & 0
\end{pmatrix}
\right] \begin{pmatrix}
\frac{\hbar^3}{2}\psi''(z)+\frac{\hbar}{4}P_2 \psi(z)\\
-\frac{\hbar^2}{\sqrt{2}} \psi'(z)\\
\hbar\psi(z)
\end{pmatrix}=0.
\end{equation}
The SL(3)-oper equation can therefore be interpreted as a flat SL(3,$\C$)-connection $\nabla$ in a jet bundle over $C$. 

On the other hand, the WKB ansatz (\ref{eqn: WKBansatz}) could be interpreted as a solution to the following first-order differential equation
\begin{equation}
\left(\partial_z-\hbar^{-1}\lambda_i^\theta(z)\right)\psi_i^\theta(z)=0.
\end{equation}
This is equivalent to say that $\psi_i^\theta(z)$ corresponds to a flat section of an abelian GL(1,$\C$)-connection $\nabla^{\text{ab},\theta}$ in a line bundle $\cL$ over the Seiberg-Witten curve $\tC$, where $i$ labels the sheet index of the covering $\tC\to C$. As described in \cite{Hollands:2013qza}, $\nabla^{\text{ab},\theta}$ has monodromy $-1$ around branch points of $\tC \to C$, therefore strictly speaking $\nabla^{\text{ab},\theta}$ is an {\it almost-flat} connection over $\tC$.

From this point of view, finding WKB solutions $\psi_i^\theta(z)$ to the SL(3)-oper equation (\ref{eqn: order3oper}), could be reformulated as finding the map from a flat SL(3,$\C$)-connections $\nabla$ over $C$ to an almost-flat GL(1,$\C$)-connection $\nabla^{\text{ab},\theta}$ over $\tC$, using the Stokes graph $W$ at phase $\theta$. Such procedure is denoted as the $W$-abelianization of $\nabla$ and has been studied in \cite{Gaiotto:2012rg,Hollands:2013qza,nikolaev2019abelianisation}. For a given $\nabla$ and a given Stokes graph $W$, there are finitely many $W$-abelianizations of $\nabla$. For the choice of $\theta=\text{arg}(\hbar)$, exact WKB analysis picks up a distinguished $W$-abelianization of $\nabla$ \cite{NeitzkeNikolaev}.

Concretely, the $W$-abelianization procedure spells out a gluing formula for local solutions across $\theta$-Stokes curves. Recall that in the complement of the $\theta$-Stokes graph, we have bases of local solutions $\psi_i^\theta$ to the SL(3)-oper equation (\ref{eqn: order3oper}), however such solutions could be different on the two sides of a $\theta$-Stokes curve. One can nevertheless give a gluing map which takes solutions on the left hand side of a $\theta$-Stokes curve to solutions on the right hand side. Across a $\theta$-Stokes curve of type $ij$, the gluing prescription could be summarized as
\begin{equation}\label{eqn:gluesinglewall}
\begin{pmatrix}
\psi_i^L\\ \psi_j^L \\ \psi_k^L
\end{pmatrix}\mapsto
\begin{pmatrix}
1 & x & 0\\
0 & 1 & 0\\
0 & 0 & 1
\end{pmatrix}
\begin{pmatrix}
\psi_i^L\\ \psi_j^L \\ \psi_k^L
\end{pmatrix}=
\begin{pmatrix}
\frac{[\psi_i^L,\psi_j^L,\psi_k^L]}{[\psi_i^R,\psi_j^L,\psi_k^L]}\psi_i^R\\
\frac{[\psi_j^L,\psi_k^L,\psi_i^L]}{[\psi_j^R,\psi_k^L,\psi_i^L]}\psi_j^R\\
\frac{[\psi_k^L,\psi_i^L,\psi_j^L]}{[\psi_k^R,\psi_i^L,\psi_j^L]}\psi_k^R
\end{pmatrix},
\end{equation} 
where $x$ is certain constant, and $[\psi_i,\psi_j,\psi_k]$ denotes the Wronskian of the three solutions.

As we will see in Sections \autoref{cha: SU3strong} and \autoref{cha: SU3weak}, for special values of $\theta$ it could happen that a $\theta$-Stokes curve of type $ij$ coincides with a $\theta$-Stokes curve of type $ji$. In this case, the gluing prescription is chosen to be\footnote{As described in \cite{Hollands:2019wbr}, there are other possible choices for the gluing formula, the one we use here amounts to an ``averaged" choice.}
\begin{equation}\label{eqn:gluedoublewall}
\begin{pmatrix}
\psi_i^L\\ \psi_j^L \\ \psi_k^L
\end{pmatrix}\mapsto
\begin{pmatrix}
z & x & 0\\
y & z & 0\\
0 & 0 & 1
\end{pmatrix}
\begin{pmatrix}
\psi_i^L\\ \psi_j^L \\ \psi_k^L
\end{pmatrix}=\begin{pmatrix}
\sqrt{\frac{[\psi_i^L,\psi_j^L,\psi_k^L] [\psi_i^L,\psi_j^R,\psi_k^L]}{[\psi_i^R,\psi_j^R,\psi_k^L][\psi_i^R,\psi_j^L,\psi_k^L]}}\psi_i^R\\
\sqrt{\frac{[\psi_j^L,\psi_i^L,\psi_k^L] [\psi_j^L,\psi_i^R,\psi_k^L]}{[\psi_j^R,\psi_i^R,\psi_k^L][\psi_j^R,\psi_i^L,\psi_k^L]}}\psi_j^R\\
\frac{[\psi_k^L,\psi_i^L,\psi_j^L]}{[\psi_k^R,\psi_i^L,\psi_j^L]}\psi_k^R
\end{pmatrix},
\end{equation}
where $z^2-xy=1$.

\subsection{The spectral coordinates}\label{sec:spectralcoordinates}

The spectral coordinates are defined as holonomies of the almost-flat GL(1,$\C$)-connection $\nabla^{\text{ab},\theta}$ along 1-cycles $\gamma$ on $\tC$:
\begin{equation}
\cX_\gamma^\theta=\text{Hol}_\gamma \nabla^{\text{ab},\theta}\in\C^\times, \quad \gamma\in H_1(\tC,\Z).
\end{equation}

By applying the gluing formulas (\ref{eqn:gluesinglewall}) and (\ref{eqn:gluedoublewall}), $\cX_\gamma^\theta$ can be expressed in terms of Wronskians of local solutions $\psi_i^\theta(z)$ to the SL(3)-oper equation (\ref{eqn: order3oper}). For the $SU(3)$ equation (\ref{eqn:SU3eqn}) considered in this notes, as we will see in Sections \autoref{cha: SU3strong} and \autoref{cha: SU3weak}, the gluing prescriptions impose relations among local solutions $\psi_i^\theta(z)$ in different regions separated by $\theta$-Stokes curves. After solving such constraints, in the end $\cX_\gamma^\theta$ are expressed in terms of Wronskians of distinguished local solutions: either as asymptotically decaying solutions as $z$ approaches a singularity, or as eigenvectors of the monodromy around a loop. In this way, $\cX_\gamma^\theta$ are identified with certain coordinate functions on a moduli space of flat SL(3,$\C$)-connections. In particular, for certain special loci in the weak-coupling region, as will be described in Section \autoref{cha: SU3weak}, we obtain coordinates of higher length-twist type generalizing complexified Fenchel-Nielsen coordinates \cite{Nekrasov:2011bc,kabaya2013parametrization}. (Higher length-twist coordinates have also been studied in other examples in \cite{Hollands:2017ahy,Jeong:2018qpc}.)

The spectral coordinates depend on the following quantities: the phase $\theta$, the Planck's constant $\hbar$ and the potentials $P_2$ and $P_3$ encoding Coulomb branch parameters etc. As long as the topology of $\theta$-Stokes graph doesn't change, the dependence on $\theta$ is trivial. However, in the $(P_2,P_3,\theta)$ parameter space there is a codimension-1 locus where the topology of $\theta$-Stokes graph changes \cite{Gaiotto:2009we,Gaiotto:2012rg}; such locus corresponds to existence of 4d BPS states and is often denoted as the BPS locus. Across the BPS locus, $\cX_\gamma^\theta$ jump by the Kontsevich-Soibelman transformation \cite{Kontsevich:2008fj,Gaiotto:2009we,Gaiotto:2012rg}.

The spectral coordinates have nice asymptotic properties: as $\hbar\to 0$ in the half plane $\mathbb{H}_\theta$ defined in (\ref{eqn:Hthetaplane}), $\cX_\gamma^\theta$ is expected to admit the following asymptotic expansion \cite{Hollands:2019wbr}:
\begin{equation}\label{eqn:asympexpansion}
\cX_\gamma^\theta\sim \text{exp}\left[\frac{1}{\hbar}\oint_\gamma\lambda^{\text{formal}}dz\right].
\end{equation}
If $\hbar$ is exactly located on the center ray of $\mathbb{H}_\theta$, or equivalently $\theta=\text{arg}(\hbar)$, then $\cX_\gamma^{\text{arg}(\hbar)}$ have stronger properties \cite{Hollands:2019wbr}. If $(P_2,P_3,\theta)$ is not on the BPS locus, $\cX_\gamma^{\text{arg}(\hbar)}$ is conjectured to be the Borel summation of the asymptotic expansion (\ref{eqn:asympexpansion}). If $(P_2,P_3,\theta)$ happens to be on the BPS locus, then the corresponding Borel transform might have singularities\footnote{This happens if $\gamma$ has non-trivial DSZ pairing with the IR charge of the 4d BPS state appearing at the BPS locus.} and the asymptotic series might not be Borel summable, in which case $\cX_\gamma^{\text{arg}(\hbar)}$ is expected to produce the median Borel summation from (\ref{eqn:asympexpansion}). In summary, asymptotic properties of the spectral coordinates enables one to compute Borel resummed quantum periods series (\ref{eqn:QPexpansion}) using abelianization methods.

\newpage
\section{The \texorpdfstring{$SU(3)$}{SU(3)} equation in the strong-coupling region}\label{cha: SU3strong}

From now on, we specialize to the $SU(3)$ equations (\ref{eqn:SU3eqn}), which corresponds to the quantum Seiberg-Witten curve for 4d $N=2$ pure $SU(3)$ SYM. In this section we focus on the analysis in the strong-coupling region, while in Section \autoref{cha: SU3weak} we study (\ref{eqn:SU3eqn}) in the weak-coupling region. 

\subsection{The BPS spectrum}
In the strong-coupling region,  $u_1$ and $u_2$ are small and the BPS spectrum is finite; it consists of 12 BPS states \cite{Alim:2011kw,Gaiotto:2012rg,Galakhov:2013oja,Cirafici:2017iju}. We choose a positive basis $\{\gamma_1,\gamma_2,\gamma_3,\gamma_4\}$ for the IR electromagnetic charge lattice. Geometrically $\gamma_i$ corresponds to homology classes in $H_1(\tC,\Z)$ where $\tC$ is the Seiberg-Witten curve; we show representative cycles in these holomogy classes in \autoref{fig:strong-Stokes}. Identifying the Dirac-Schwinger-Zwanziger pairing with the intersection pairing in $H_1(\tC,\Z)$, we obtain the following pairing matrix with respect to the chosen basis $\{\gamma_1,\gamma_2,\gamma_3,\gamma_4\}$:
\begin{equation}
\begin{pmatrix}
0 & 0 & 1 &-2\\
0 & 0 & -2 & 1\\
-1 & 2 & 0 & 0\\
2 & -1 & 0 & 0
\end{pmatrix}.
\end{equation}
The 12 BPS states have the following IR charges:
\begin{equation}
\pm \gamma_1, \pm \gamma_2,\pm\gamma_3,\pm\gamma_4,\pm (\gamma_1+\gamma_3),\pm(\gamma_2+\gamma_4).
\end{equation}

The BPS spectrum respects a $\Z_6$ symmetry induced by the action of discrete R-symmetry on the central charge operator. In particular if we take $u_1=u_2=0$, the BPS states have phases $0,\pi/3,2\pi/3,\pi,4\pi/3,5\pi/3$ and their central charges have the same norm. This is illustrated in \autoref{fig:Strong-BPS}.
\insfigsvg{Strong-BPS}{0.5}{Central charges of the 12 BPS states at $u_1=u_2=0$.}

\subsection{Stokes graphs}

As described in Section \autoref{sec:WKBabelianization}, the Stokes graph appearing in the exact WKB method depends on a phase parameter $\theta$. In particular if $\theta$ happens to be a phase equal to the central charge phase of a 4d BPS hypermultiplet, the corresponding Stokes graph contains finite segments or webs; while if $\theta$ happens to be equal to the phase of a 4d BPS vector multiplet, the Stokes graph would contain annulus domain. Such phases are usually called {\it critical phases}. For example, taking $u_1= 0.3, u_2=0$ and $\Lambda=1$, in \autoref{fig:smallu1} we show the Stokes graphs zoomed in around $z=0$, at critical phases $\theta\approx0.3037\pi,0.6963\pi,\pi$.\footnote{The Stokes graph at phase $\theta+\pi$ looks almost identical to that at phase $\theta$, except the arrow directions are reversed. Therefore it suffices to consider Stokes graphs at phases within $(0,\pi]$.} (Details of the Stokes graphs at large $|z|$ are similar to what is shown in \autoref{fig:strong-Stokes}.)
\insfigpng{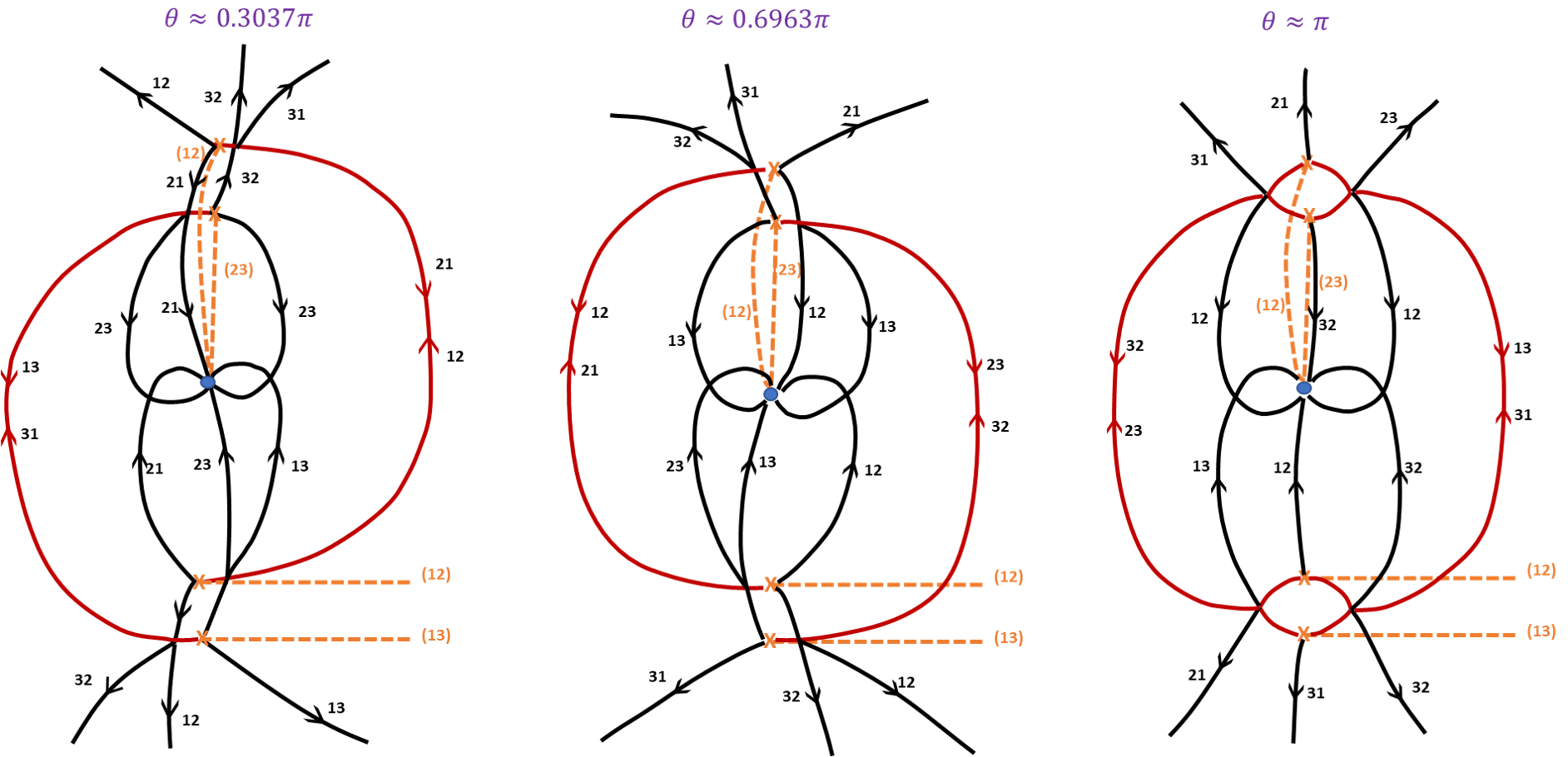}{0.5}{Zoomed-in Stokes graphs at $\theta\approx0.3037\pi,0.6963\pi,\pi$, where we have taken $u_1=0.3, u_2=0, \Lambda=1$ and we focus on details near $z=0$. The irregular singularity at $z=0$ is denoted by a blue dot, the irregular singularity at $z=\infty$ is not shown in the figure. Branch points are represented as orange crosses, while orange dashed lines denote the choice of branch cuts. Finite segments or webs are colored in red; in class $S$ setup \cite{Gaiotto:2009hg,Gaiotto:2009we,Gaiotto:2012rg} they correspond to trajectories of 6d BPS strings which give rise to 4d BPS states after the twisted compactification on $C$.}

At the special point $u_1=u_2=0$, Stokes graphs at critical phases all look the same, except that the sheet labeling gets permuted. As an example, \autoref{fig:strong-Stokes} shows the Stokes graph at $\theta=\pi/3$. In the following, we will derive an expression for the spectral coordinates in terms of distinguished local solutions to the $SU(3)$ equation (\ref{eqn:SU3eqn}). 
\insfigpng{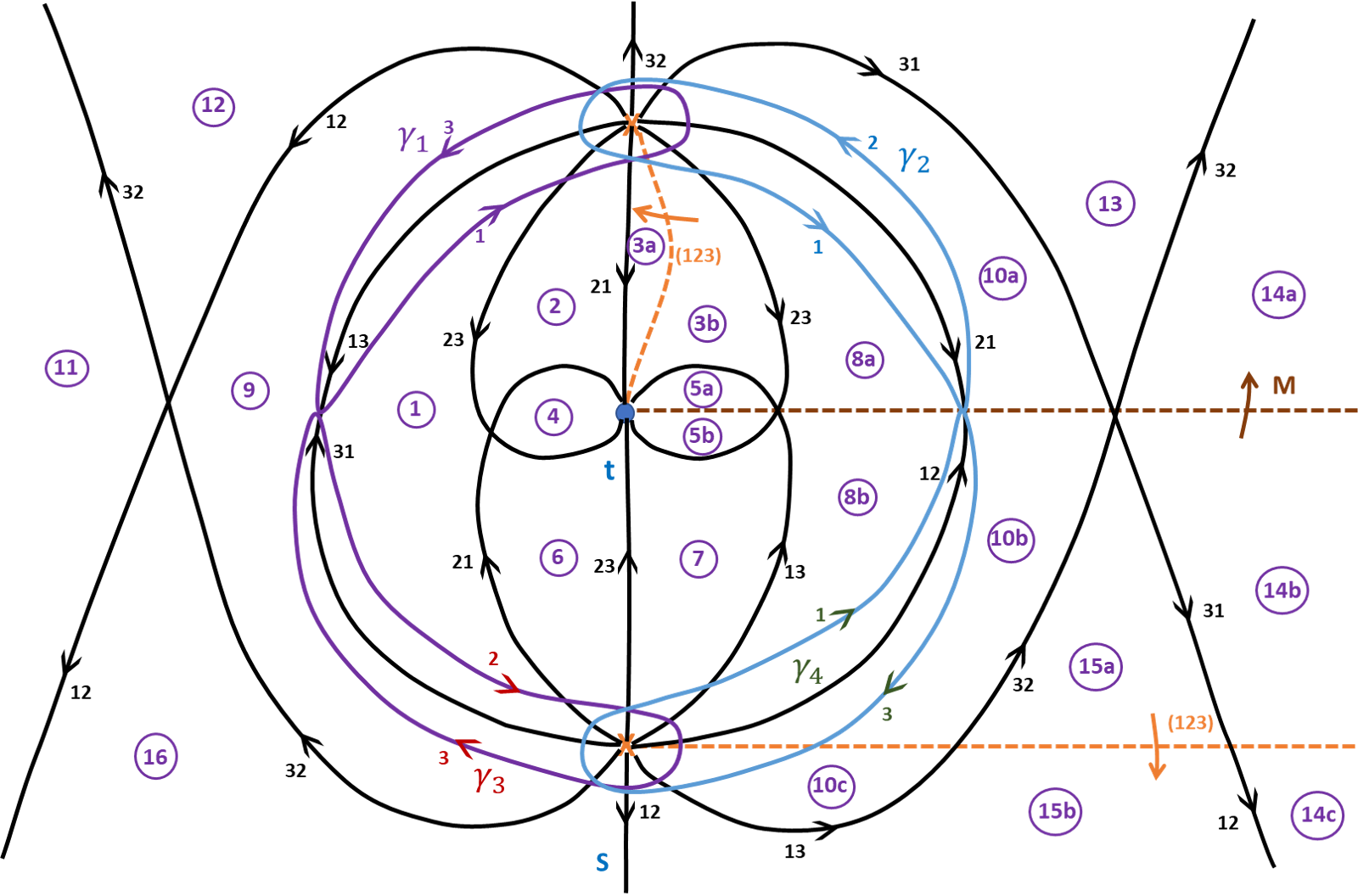}{0.5}{The Stokes graph at $\theta=\pi/3$ with $u_1=u_2=0$ and $\Lambda=1$. The irregular singularity at $z=0$ is represented by a blue dot, while the irregular singularity at $z=\infty$ is not shown in the figure. Branch points and branch cuts are denoted by orange crosses and orange dashed lines respectively. The monodromy cut is represented by a brown dotted line. We also show representative cycles corresponding to the basis charges $\gamma_1$, $\gamma_2$, $\gamma_3$ and $\gamma_4$ in purple, blue, red and green respectively. The Stokes graph divides the puntured-plane into 16 regions, which are labeled by circled purple numbers.}

\subsection{Solving the abelianization problem}{\label{sec:strongabsol}}

The Stokes graph in \autoref{fig:strong-Stokes} divides the punctured-plane into 16 regions; in each region we could choose a basis of local WKB solutions to the SU(3) equation (\ref{eqn:SU3eqn}). In particular there are two distinguished local solutions:
\begin{itemize}
	\item The local WKB solution $t$ near $z=0$, characterized as a solution which decays exponentially as $z\to 0$ along the negative imaginary axis.
	
	\item The local WKB solution $s$ near $z=\infty$, characterized as a solution which decays exponentially as $z\to \infty$ along the negative imaginary axis.
\end{itemize}

Based on the gluing rules in Section \autoref{sec:gluingrules}, we could immediately identify a few local WKB solutions in some regions in \autoref{fig:strong-Stokes} as $s$, $t$ or their images under the monodromy actions $M$ and $M^{-1}$ around the irregular singularity. However, across the 16 regions there are 10 local WKB solutions which are not straightforwardly related to the distinuighed ones; we denote these solutions as $\psi_i~ (i=1,...,10)$. The local WKB solutions in the 16 regions are listed in Table \ref{table:strong}. Here to write the basis concretely as an ordered tuple we have used the trivialization of $\tC$ away from the branch cuts.
\begin{table}[h]
	\centering
	\begin{tabular}{|c|c|c|c|} \hline
		region   & basis & region             & basis \\
		\hline 
		$1$    & $\big(Mt,s,t\big)$   & $2$                & $\big(Mt,\psi_1,t\big)$\\
		\hline
		$3a$    & $\big(Mt,\psi_2,t\big)$   & $3b$                & $\big(\psi_2,t,Mt\big)$\\
		\hline
		$4$    & $\big(Mt,\psi_3,t\big)$   & $5a$                & $\big(M^2t,t,Mt\big)$\\
		\hline
		$5b$    & $\big(Mt,M^{-1}t,t\big)$   & $6$                & $\big(Mt,\psi_4,t\big)$\\
		\hline
		$7$    & $\big(Mt,\psi_5,t\big)$   & $8a$                & $\big(\psi_2,M\psi_5,Mt\big)$\\
		\hline
		$8b$    & $\big(M^{-1}\psi_2,\psi_5,t\big)$   & $9$                & $\big(\psi_6,s,\psi_7\big)$\\
		\hline
		$10a$    & $\big(Ms,s,Mt\big)$   & $10b$                & $\big(s,M^{-1}s,t\big)$\\
		\hline
		$10c$    & $\big(t,s,M^{-1}s\big)$   & $11$                & $\big(Ms,s,M^{-1}s\big)$\\
		\hline
		$12$    & $\big(Ms,s,\psi_7\big)$   & $13$                & $\big(Ms,s,\psi_8\big)$\\
		\hline
		$14a$ & $\big(Ms,s,\psi_9\big)$ & $14b$  & $\big(s,M^{-1}s,M^{-1}\psi_9\big)$\\
		\hline
		   $14c$ & $\big(M^{-1}\psi_9,s,M^{-1}s\big)$	& $15a$  & $\big(s,M^{-1}s,\psi_{10}\big)$  \\
		\hline
	 $15b$ & $\big(\psi_{10},s,M^{-1}s\big)$ & $16$  & $\big(\psi_6,s,M^{-1}s\big)$\\
		\hline
	\end{tabular}
\caption{Bases of local WKB solutions for the 16 regions shown in \autoref{fig:strong-Stokes}.}
\label{table:strong}
\end{table}

The concrete task of abelianization here, is solving $\psi_i$ in terms of distinguished local solutions and the monodromy $M$. For this purpose, we consider constraints coming from gluing factors across $\theta$-Stokes curves. As an example, the gluing factor across the $\theta$-Stokes curve of type $13$ between regions $3b$ and $5a$ indicates that $\psi_2, Mt, M^2t$ are co-planar in the solution space. Let us denote $\lr{\alpha}{\beta}$ as the two-dimensional plane spanned by $\alpha$ and $\beta$ in the three-dimensional space of local solutions, then we have $\langle\psi_2,Mt\rangle=\langle M^2t,Mt\rangle$. Similarly, the gluing factor across the coincident $\theta$-Stokes curves of type $12$ and $21$ between regions $8a$ and $10a$ implies $\langle\psi_2,M\psi_5\rangle=\langle Ms,s \rangle$. For generic $M$, $\psi_2$ then corresponds to the intersection between $\lr{Ms}{s}$ and $\lr{M^2t}{Mt}$. All the 10 local WKB solutions $\psi_i$ could be determined in a similar fashion:
\begin{equation}{\label{eqn:strongsol}}
\begin{split}
&\psi_1=\lr{s}{t}\cap \lr{M^2t}{Mt},\quad 
\psi_2=\lr{Ms}{s}\cap\lr{M^2t}{Mt},\\
&\psi_3=\lr{M^2t}{Mt}\cap\lr{M^{-1}t}{t},\quad
\psi_4=\lr{Mt}{s}\cap\lr{M^{-1}t}{t},\\
&\psi_5=\lr{s}{M^{-1}s}\cap\lr{M^{-1}t}{t},\quad
\psi_6=\lr{Ms}{s}\cap\lr{Mt}{t},\\
&\psi_7=\lr{M^{-1}s}{s}\cap\lr{Mt}{t},\quad
\psi_8=\lr{M^{-1}s}{s}\cap\lr{Mt}{Ms},\\
&\psi_9=\lr{M^{-1}s}{s}\cap\lr{M^2s}{Ms},\quad
\psi_{10}=\lr{t}{M^{-1}s}\cap\lr{Ms}{s}.
\end{split}
\end{equation}

\subsection{The spectral coordinates}

Let $\cX_{\gamma_i}$\footnote{From now on, we suppress the dependence on $\theta$ in the notation for spectral coordinates for compactness reasons.} denote the holonomy of $\nabla^{\text{ab},\theta}$ along cycles $\gamma_i$ shown in \autoref{fig:strong-Stokes}; $\cX_{\gamma_i}$ are then certain spectral coordinates for the flat SL(3,$\C$)-connection $\nabla$. We remark that, although the Stokes graph in \autoref{fig:strong-Stokes} was drawn at the special point $u_1=u_2=0$, the expressions for $\cX_{\gamma_i}$ in terms of distinguished local solutions hold as we move to small non-zero $u_{1,2}$; the Stokes graph doesn't go through topological changes in that process.

Applying the gluing formulas (\ref{eqn:gluesinglewall}) and (\ref{eqn:gluedoublewall}), these spectral coordinates are given in terms of Wronskians of distinguished local solutions:
\begin{equation}\label{eqn:Xstrong}
\begin{split}
\cX_{\gamma_1}&=\frac{[Mt,Ms,s][\psi_7,\psi_6,s][t,M^{-1}s,s]}{[\psi_7,s,Ms][t,Mt,s][\psi_6,s,M^{-1}s]},\\
\cX_{\gamma_2}&=\frac{[s,t,Mt][M^{-1}\psi_2,\psi_5,t][s,Mt,t]}{[M^{-1}\psi_2,t,M^{-1}t][Ms,s,Mt][\psi_5,t,Mt]},\\
\cX_{\gamma_3}&=\frac{[\psi_5,s,t][M^{-1}s,s,\psi_6][M^{-1}t,t,M^{-1}\psi_2]}{[\psi_5,t,Mt][M^{-1}s,s,t][M^{-1}s,M^{-1}\psi_2,t]}\sqrt{\frac{[s,Mt,t][\psi_7,Mt,s]}{[\psi_7,s,\psi_6][t,\psi_6,s]}},\\
\cX_{\gamma_4}&=\frac{[\psi_6,t,s][Mt,t,\psi_5],[Ms,s,\psi_7]}{[\psi_6,s,M^{-1}s][Mt,t,s][Mt,\psi_7,s]}\sqrt{\frac{[t,M^{-1}s,s][M^{-1}\psi_2,M^{-1}s,t]}{[M^{-1}\psi_2,t,\psi_5][s,\psi_5,t]}},
\end{split}
\end{equation}
where $\psi_{2,5,6,7}$ are intersections of certain planes spanned by the distinguished local solutions $s,t$ and their images under the monodromy action; they are given explicitly in (\ref{eqn:strongsol}). 

As a remark, the expressions for $\cX_{\gamma_i}$ in (\ref{eqn:Xstrong}) correspond to the critical Stokes graph (\autoref{fig:strong-Stokes}) at $\theta=\pi/3$. Following the same procedure as described above, we also obtain expressions for the spectral coordinates $\cX_{\gamma_i}^\pm$ at $\theta=\pi/3+\epsilon$ and $\theta=\pi/3-\epsilon$ for small $\epsilon$. $\cX_{\gamma_{1}}^\pm$ and $\cX_{\gamma_2}^\pm$ have the same expression as in (\ref{eqn:Xstrong}). Meanwhile $\cX_{\gamma_3}^+$ ($\cX_{\gamma_4}^+$) differs from $\cX_{\gamma_3}^-$ ($\cX_{\gamma_4}^-$) by a Kontsevich-Soibelman transformation involving $\cX_{\gamma_{1,2}}$. The $\cX_{\gamma_{3,4}}$ in (\ref{eqn:Xstrong}) could be thought of as an average of $\cX_{\gamma_{3,4}}^\pm$. These properties are consistent with the expected Stokes phenomena for the Borel-resummed quantum periods, which are controlled by the BPS spectrum illustrated in \autoref{fig:Strong-BPS}. Concretely across the phase $\theta=\pi/3$, spectral coordinates $\cX_\gamma$ where $\gamma$ has non-trivial DSZ pairing with $\gamma_{1,2}$ would jump by a Kontsevich-Soibelman transformation.

We further remark that the Wronskian expressions of $\cX_\gamma$ in (\ref{eqn:Xstrong}) are the same for both quantization choices in (\ref{eqn:SU3eqn}); this is because the Stokes graph is determined by the classical Seiberg-Witten curve, as described in \autoref{sec:WKBsol}. However, it is true that the concrete special solutions such as $s$ and $t$ differ between the two quantization choices, resulting different numerical values of $\cX_\gamma$.

\subsection{The asymptotic behavior}\label{sec:strongsecasymp}

As $\hbar\to 0$ while staying within the half-plane $\mathbb{H}_\theta$ defined in (\ref{eqn:Hthetaplane}), $\hbar\text{log}\cX_{\gamma}$ is conjectured to have an asymptotic series expansion given by the formal quantum periods series (\ref{eqn:QPexpansion}). We could perform numerical checks against this conjecture. On the one hand, we evaluate (\ref{eqn:Xstrong}) by numerically solving (\ref{eqn:SU3eqn}) and computing Wronskians of distinguished solutions. On the other hand, we can compute the truncated series in $\hbar$ of the formal quantum periods $\Pi_\gamma(\hbar)$. For example we first consider the differential equation $A$ in (\ref{eqn:SU3eqn}), taking $u_1=u_2=0$ and $\Lambda=1$, numerical results for $\hbar=\text{e}^{\text{i}\pi/3}$ and $\hbar=\frac{1}{2}\text{e}^{\text{i}\pi/3}$ are listed in Table \ref{table:asympstrong}.

\begin{table}[h]
	\centering
	\begin{tabular}{|c|c|c|c|c|}
		\hline
		& \multicolumn{2}{|c|}{$\hbar=\text{e}^{\text{i}\pi/3}$} & \multicolumn{2}{|c|}{$\hbar=\frac{1}{2}\text{e}^{\text{i}\pi/3}$} \\
		\hline 
	    & evaluation of (\ref{eqn:Xstrong}) & $\frac{1}{\hbar}\Pi_\gamma(\hbar)$ at $o(\hbar^6)$ & evaluation of (\ref{eqn:Xstrong}) & $\frac{1}{\hbar}\Pi_\gamma(\hbar)$ at $o(\hbar^6)$\\
		\hline
		$\text{log}\cX_{\gamma_1}$&  $-3.5756$ & $-3.5873$ & $-10.2481$& $-10.2482$ \\
		\hline
		$\text{log}\cX_{\gamma_2}$&  $-3.5756$ & $-3.5873$ & $-10.2481$ & $-10.2482$\\
		\hline
		$\text{log}\cX_{\gamma_3}$& $1.7878-0.4490\text{i}$ & $1.7937-0.4405\text{i}$& $5.12405+1.88253\text{i}$ & $5.12412+1.88266\text{i}$\\
		\hline
		$\text{log}\cX_{\gamma_4}$ & $1.7878-0.4490\text{i}$ & $1.7937-0.4405\text{i}$& $5.12405+1.88253\text{i}$& $5.12412+1.88266\text{i}$\\
		\hline
	\end{tabular}
\caption{Comparison of $\text{log}\cX_\gamma$ with the formal quantum periods expansion up to order-$\hbar^6$ at $\hbar=\text{e}^{\text{i}\pi/3}$ and $\hbar=\frac{1}{2}\text{e}^{\text{i}\pi/3}$ for equation $A$ of (\ref{eqn:SU3eqn}), where we have set $u_1=u_2=0$ and $\Lambda=1$.}
\label{table:asympstrong}
\end{table}

From \autoref{table:asympstrong} we see that as $|\hbar|$ gets smaller, the spectral coordinates obtained via abelianization method get closer to the truncated quantum periods expansion\footnote{As $\Pi_\gamma(\hbar)$ is an asymptotic series, there exist an optimal term where the agreement would be the best. Here we approximate the optimal term by numerical experiments}. This is in line with the conjectured asymptotic behavior of spectral coordinates. Unfortunately numerical evaluation of (\ref{eqn:Xstrong}) becomes rather difficult for small $|\hbar|$; we are not able to demonstrate the asymptotic behavior at smaller $|\hbar|$. Nevertheless we regard Table \ref{table:asympstrong} as evidence that higher-order exact WKB methods indeed works. As a further remark, we notice that the system has a symmetry where $\cX_{\gamma_1}=\cX_{\gamma_2}$ and $\cX_{\gamma_3}=\cX_{\gamma_4}$. The numerical values in \autoref{table:asympstrong} do respect such symmetry, which could serve as an extra consistency check for our analysis. We perform similar analysis for equation $B$ of (\ref{eqn:SU3eqn}). Here the numerical agreement seems slightly better; we list the values at $\hbar=\frac{1}{2}\text{e}^{\text{i}\pi/3}$ in Table \ref{table:asympstrong2}. The asymptotic analysis suggests that both differential equations $A$ and $B$ of (\ref{eqn:SU3eqn}) are valid quantization choices of the classical Seiberg-Witten curve, although the differential equation $B$ is the canonical quantization from the gauge theory point of view, as we will describe in upcoming \cite{Yan2}.

\begin{table}[h]
	\centering
	\begin{tabular}{|c|c|c|}
		\hline
		&  \multicolumn{2}{|c|}{$\hbar=\frac{1}{2}\text{e}^{\text{i}\pi/3}$} \\
		\hline 
		& evaluation of (\ref{eqn:Xstrong}) & $\frac{1}{\hbar}\Pi_\gamma(\hbar)$ at $o(\hbar^6)$ \\
		\hline
		$\text{log}\cX_{\gamma_1}$&  $-11.21119$ & $-11.21120$  \\
		\hline
		$\text{log}\cX_{\gamma_2}$&  $-11.21119$ & $-11.21120$ \\
		\hline
		$\text{log}\cX_{\gamma_3}$& $5.60559 + 2.71805\text{i}$ & $5.60560 + 2.71808\text{i}$\\
		\hline
		$\text{log}\cX_{\gamma_4}$ & $5.60559 + 2.71805\text{i}$ & $5.60560 + 2.71808\text{i}$\\
		\hline
	\end{tabular}
	\caption{Comparison of $\text{log}\cX_\gamma$ with the formal quantum periods expansion up to order-$\hbar^6$ at $\hbar=\frac{1}{2}\text{e}^{\text{i}\pi/3}$ for equation $B$ of (\ref{eqn:SU3eqn}), where we have set $u_1=u_2=0$ and $\Lambda=1$.}
	\label{table:asympstrong2}
\end{table}

\newpage
\section{The \texorpdfstring{$SU(3)$}{SU(3)} equation in the weak-coupling region}\label{cha: SU3weak}

\subsection{A Stokes graph}

\insfigpng{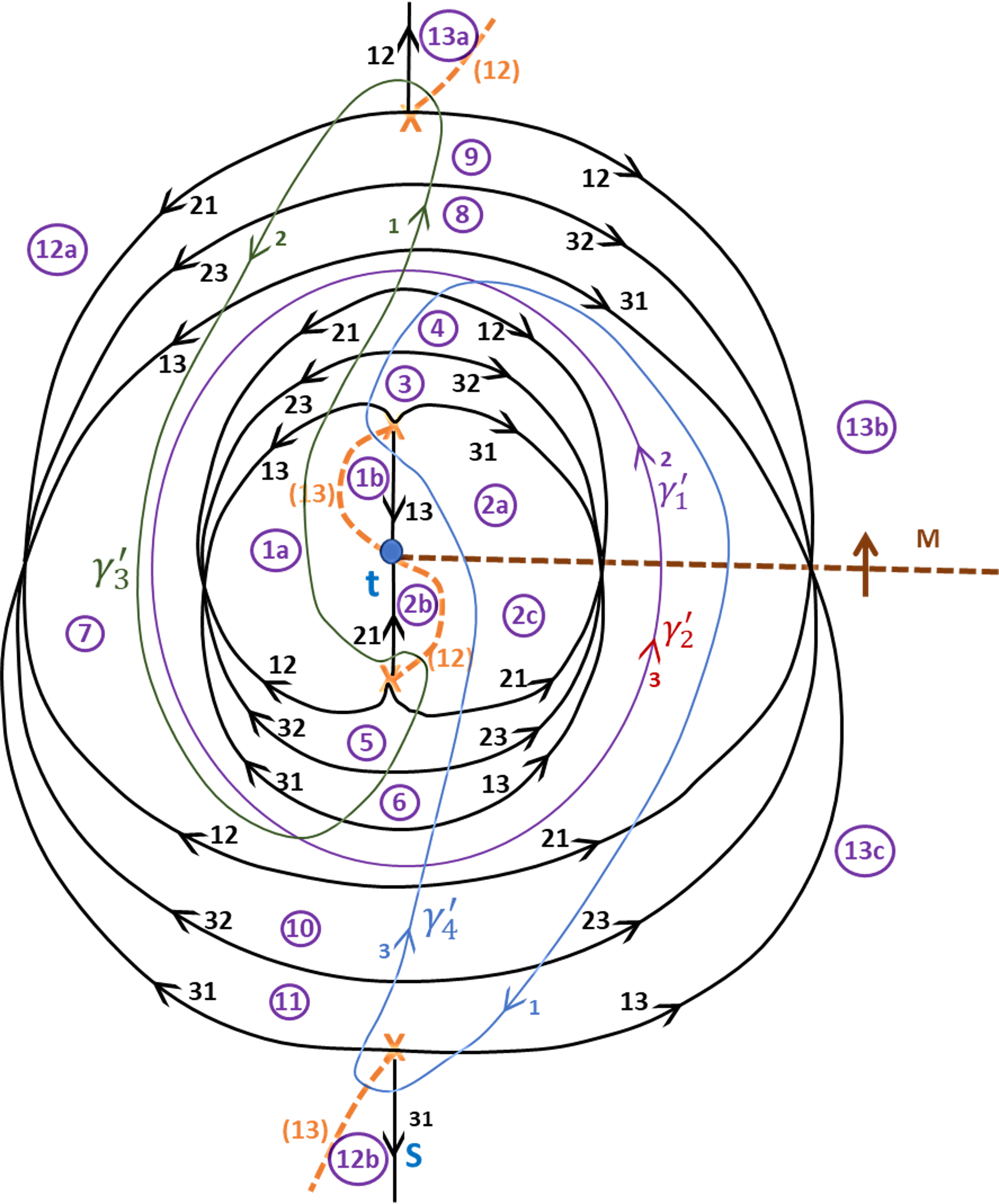}{0.21}{The Stokes graph at $\theta=0$, with $u_1=4.5$, $u_2=0$ and $\Lambda=1$. The notation conventions are the same as those in \autoref{fig:strong-Stokes}. We show representative cycles corresponding to the basis charges $\gamma'_1$, $\gamma'_2$, $\gamma'_3$ and $\gamma'_4$ in purple, red, green and blue respectively. This Stokes graph divides the punctured-plane into 13 regions.}

Now we turn our attention to the weak-coupling region, where the BPS spectrum becomes very complicated \cite{Galakhov:2013oja}. In particular we consider parameters under which a Stokes graph of the higher length-twist type \cite{Hollands:2017ahy} appears. Our motivation for considering such Stokes graphs is, as briefly described in Section \autoref{sec:Instantonmethod}, to make connections with gauge theory calculations. The corresponding spectral coordinates are higher length-twist coordinates, which are closely related to the quantum $a$ and $a_D$ periods. In this paper we construct the higher length-twist coordinates using abelianization methods, while perspectives from instanton calculus is investigated in \cite{Yan2}. 

Taking $\Lambda=1$, $u_2=0$, and real $u_1>1.25$, at $\theta=0$ the Stokes graph looks like \autoref{fig:weak-Stokes}. In particular we see a ring domain corresponding to the $W$-bosons; this ring domain is denoted as the region 7 in \autoref{fig:weak-Stokes}. We choose a basis $\{\gamma'_1,\gamma'_2,\gamma'_3,\gamma'_4\}$ of the IR charge lattice, as shown in \autoref{fig:weak-Stokes}. This basis is more natural for the analysis in the weak-coupling region. We will write down the corresponding spectral coordinates $\cX_{\gamma'_i}$ in Section \autoref{sec:SCweak}; in particular $\cX_{\gamma'_{1,2}}$ are higher length coordinates and $\cX_{\gamma'_{3,4}}$ are higher twist coordinates.

\subsection{Solving the abelianization problem}
The Stokes graph in \autoref{fig:weak-Stokes} divides the punctured-plane into 13 regions. The local WKB solutions in each region turn out to be expressed in terms of the following distinguished local solutions:
\begin{itemize}
\item The local WKB solution $t$ near $z=0$ and the local WKB solution $s$ near $z=\infty$, characterized as exponentially decaying solution as $z\to0$ and $z\to\infty$ respectively, along the negative imaginary axis.

\item Eigenvectors $\alpha$, $\beta$ and $\gamma$ of the counterclockwise monodromy $M$ around $z=0$. For the parameter range we consider here, namely real $u_1>1.25$, $u_2=0$ and $\hbar\in\R_+$, the eigenvalues of the monodromy $M$ are all real and positive, moreover one of the eigenvalues is 1. Let us denote the eigenvalues of $M$ as $\mu_\alpha$, $\mu_\beta$ and $\mu_\gamma$, corresponding to the eigenvectors $\alpha$, $\beta$ and $\gamma$ respectively. Then $\alpha$, $\beta$ and $\gamma$ are specified according to the condition $\mu_\alpha=1$, $\mu_\beta>1$ and $\mu_\gamma<1$.\footnote{One might wonder where this particular specification for $\alpha$, $\beta$ and $\gamma$ comes from; after all, given a generic monodromy matrix $M$ we have 6 ways of defining $(\alpha,\beta,\gamma)$. The choice we take here is the one which matches the leading asymptotic behavior of spectral coordinates, given by the classical periods around the ring domain.}
\end{itemize}

We proceed by first listing bases of local WKB solutions for each region in Table \ref{table:weak}; in this case there are 14 local solutions $\phi_i$ not straightforwardly related to the above distinguished solutions and their images under the monodromy action.

\begin{table}[h]
	\centering
	\begin{tabular}{|c|c|c|c|} \hline
		region   & basis & region             & basis \\
		\hline 
		$1a$    & $\big(t,Mt,M^{-1}t\big)$   & $1b$ & $\big(M^{-1}t,Mt,t\big)$\\
		\hline
		$2a$ & $\big(M\phi_1, Mt,t\big)$ & $2b$ & $\big(t, \phi_1, M^{-1}t\big)$\\
		\hline
		$2c$ & $\big(\phi_1, t, M^{-1}t\big)$ & $3$ & $\big(\phi_2, Mt, \phi_3\big)$\\
		\hline
		$4$ & $\big(\phi_2, \phi_4, \gamma\big)$ & $5$ & $\big(\phi_5, \phi_6, M^{-1}t\big)$\\
		\hline
		$6$ & $\big(\phi_5,\beta,\phi_7\big)$ & $7$ & $\big(\alpha, \beta, \gamma\big)$\\
		\hline
		$8$ & $\big(\phi_8, \beta,\phi_9\big)$ & $9$ & $\big(\phi_8, \phi_{10}, s\big)$\\
		\hline
		$10$ & $\big(\phi_{11}, \phi_{12},\gamma\big)$ & $11$ & $\big(\phi_{11}, Ms, \phi_{13}\big)$\\
		\hline
		$12a$ & $\big(\phi_{14}, Ms,s\big)$ & $12b$ & $\big(s, Ms, \phi_{14}\big)$\\
		\hline
		$13a$ & $\big(M^2s, Ms,s\big)$ & $13b$ & $\big(Ms, M^2s, s\big)$\\
		\hline
		$13c$ & $\big(s, Ms,M^{-1}s\big)$ &  & \\
		\hline
	\end{tabular}
	\caption{Bases of local WKB solutions for the 13 regions shown in \autoref{fig:weak-Stokes}.}
	\label{table:weak}
\end{table}

Similar to the strong coupling case described in Section \autoref{sec:strongabsol}, by exploring the constraints imposed by the gluing condition we could solve for $\phi_i$. The solutions are given as follows:\begin{equation}{\label{eqn:weaksol}}
\begin{split}
&\phi_1=\lr{t}{Mt}\cap \lr{M^{-2}t}{M^{-1}t},\quad 
\phi_2=\lr{t}{M^{-1}t}\cap\lr{\alpha}{\beta},\\
&\phi_3=\lr{t}{M^{-1}t}\cap\lr{Mt}{\gamma},\quad
\phi_4=\lr{Mt}{\gamma}\cap\lr{\alpha}{\beta},\\
&\phi_5=\lr{t}{Mt}\cap\lr{\alpha}{\gamma},\quad
\phi_6=\lr{t}{Mt}\cap\lr{M^{-1}t}{\beta},\\
&\phi_7=\lr{M^{-1}t}{\beta}\cap\lr{\alpha}{\gamma},\quad
\phi_8=\lr{\alpha}{\gamma}\cap\lr{Ms}{M^2s},\\
&\phi_9=\lr{\alpha}{\gamma}\cap\lr{\beta}{s},\quad
\phi_{10}=\lr{\beta}{s}\cap\lr{Ms}{M^2s},\\
&\phi_{11}=\lr{\alpha}{\beta}\cap\lr{s}{M^{-1}s},\quad
\phi_{12}=\lr{\alpha}{\beta}\cap\lr{Ms}{\gamma},\\
&\phi_{13}=\lr{Ms}{\gamma}\cap\lr{s}{M^{-1}s},\quad
\phi_{14}=\lr{s}{M^{-1}s}\cap\lr{M^2s}{Ms}.
\end{split}
\end{equation}

\subsection{The spectral coordinates}\label{sec:SCweak}

The spectral coordinates $\cX_{\gamma'_{1,2}}$ could be identified with the higher length coordinates on the moduli space of flat SL(3,$\C$)-connections; they are simply given by the corresponding eigenvalues of the monodromy action:
\begin{equation}\label{eqn:Xweak1}
\cX_{\gamma'_1}=\mu_\beta, \quad \cX_{\gamma'_2}=\mu_\gamma
\end{equation}

The spectral coordinates $\cX_{\gamma'_{3,4}}$ are higher twist coordinates; we express them via Wronskians of distinguished local solutions:
\begin{equation}\label{eqn:Xweak2}
\begin{split}
\cX_{\gamma'_3}=&\sqrt{-\frac{[Ms,\phi_{14},s][Ms,\phi_8,s][\phi_{10},\phi_9,\phi_8][\beta,\phi_7,\phi_5][\beta,M^{-1}t,\phi_5][\phi_6,\phi_1,M^{-1}t][t,M^{-1}t,Mt]}{[\phi_{10},\phi_{14},s][\beta,\phi_9,\phi_8][\beta,s,\phi_8][\phi_6,\phi_7,\phi_5][t,\phi_1,M^{-1}t][t,\phi_5,M^{-1}t][\phi_2,\phi_3,Mt]}}\\
&\times\sqrt{\frac{[t,\phi_3,Mt][\phi_2,\phi_4,\gamma][\phi_2,\beta,\gamma][\alpha,\phi_9,\beta][\phi_8,\phi_{10},s][\phi_8,M^2s,s]}{[\phi_2,M^{-1}t,Mt][\alpha,\phi_4,\gamma][\phi_8,\phi_9,\beta][\phi_8,\gamma,\beta][Ms,M^2s,s][Ms,\phi_{10},s]}},\\
\cX_{\gamma'_4}=&\frac{1}{\mu_\alpha}\sqrt{\frac{[s,\phi_{14},Ms][s,\phi_{11},Ms][\phi_{13},\phi_{12},\phi_{11}][\gamma,\alpha,\beta][\gamma,\phi_5,\beta][\phi_7,\phi_6,\phi_5][t,M^{-1}t,Mt]}{[\phi_{13},\phi_{14},Ms][\gamma,\phi_{12},\phi_{11}][\gamma,Ms,\phi_{11}][\phi_7,\alpha,\beta][M^{-1}t,\phi_6,\phi_5][M^{-1}t,\beta,\phi_5][\phi_2,\phi_3,Mt]}}\\
&\times\sqrt{\frac{[t,\phi_3,Mt][\phi_2,\phi_4,\gamma][\phi_2,\beta,\gamma][\alpha,\phi_{12},\gamma][\phi_{11},\phi_{13},Ms][\phi_{11},M^{-1}s,Ms]}{[\phi_2,M^{-1}t,Mt][\alpha,\phi_4,\gamma][\phi_{11},\psi_{12},\gamma][\phi_{11},\beta,\gamma][s,M^{-1}s,Ms][s,\phi_{13},Ms]}},
\end{split}
\end{equation}
where the local WKB solutions $\phi_i$ are given in (\ref{eqn:weaksol}).

The expressions in (\ref{eqn:Xweak2}) look rather formidable. As a consistency check and an application, we consider the exact quantization condition (EQC) for a spectral problem relevant to the $SU(3)$ equation (\ref{eqn:SU3eqn}). Imagine we would like to study certain bound state solution to (\ref{eqn:SU3eqn}) living along a one-dimensional path between $z=0$ and $z=\infty$, where the solution decays as $z$ approaches $0$ and $\infty$. For example suppose the one-dimensional path goes into $z=0$ and $z=\infty$ along the negative imaginary axis, then the condition for existence of such bound states is that the distinguished solution $s$ decaying into $z=0$ is proportional to the distinguished solution $t$ decaying into $z=\infty$, after analytical continuation to a common region. Substituting this condition into (\ref{eqn:Xweak2}), the complicated-looking expressions simplify to 
\begin{equation}\label{eqn:EQC}
\frac{1}{\mu_\beta}\cX_{\gamma'_3}=1,\quad \mu_\beta \cX_{\gamma'_4}=1.
\end{equation}
Equation (\ref{eqn:EQC}) is regarded as the EQC for such bound states. 

In \cite{Grassi:2018bci} the authors studied EQC for a family of exactly solvable deformed Hamiltonians, obtained by quantizing the Seiberg-Witten curve for 4d $N=2$ $SU(N)$ SYM in the hyperelliptic form. In the case of $SU(3)$ SYM, the Seiberg-Witten curve we use here comes from class $S$ construction; it is in the dual parametrization of the Seiberg-Witten curve considered in \cite{Grassi:2018bci}. It would be interesting to straighten out the relation between (\ref{eqn:EQC}) and the EQC in \cite{Grassi:2018bci} derived from certain 4d limit of TS/ST correspondence.


\subsection{The asymptotic behavior}

Similar to Section \autoref{sec:strongsecasymp}, here we perform numerical checks against the asymptotic behavior of spectral coordinates. We take $u_1=4.5$, $u_2=0$ and $\Lambda=1$, some numerical results for equation $A$ and $B$ of (\ref{eqn:SU3eqn}) are listed in \autoref{table:asympweak1} and \autoref{table:asympweak2} respectively.

\begin{table}[h]
	\centering
	\begin{tabular}{|c|c|c|c|c|}
		\hline
		& \multicolumn{2}{|c|}{$\hbar=1$} & \multicolumn{2}{|c|}{$\hbar=0.7$} \\
		\hline 
		& evaluation (4.2-4.3) & $\frac{1}{\hbar}\Pi_{\gamma'}(\hbar)$ at $o(\hbar^2)$ & evaluation (4.2-4.3) & $\frac{1}{\hbar}\Pi_{\gamma'}(\hbar)$ at $o(\hbar^2)$\\
		\hline
		$\text{log}\cX_{\gamma'_1}$&  $11.90$ & $11.98$ & $18.14$& $18.17$ \\
		\hline
		$\text{log}\cX_{\gamma'_2}$&  $-11.90$ & $-11.98$ & $-18.14$ & $-18.17$\\
		\hline
		$\text{log}\cX_{\gamma'_3}$& $5.93+0.94\text{i}$ & $5.99+0.96\text{i}$& $9.080+ 2.692\text{i}$ & $9.086 + 2.697\text{i}$\\
		\hline
		$\text{log}\cX_{\gamma'_4}$ & $-5.93-0.94\text{i}$  & $-5.99-0.96\text{i}$& $-9.080- 2.692\text{i}$& $-9.086 - 2.697\text{i}$\\
		\hline
	\end{tabular}
	\caption{Comparison of $\text{log}\cX_{\gamma'}$ with the formal quantum periods expansion up to order-$\hbar^2$ for equation $A$ of (\ref{eqn:SU3eqn}) at $\hbar=1$ and $\hbar=0.7$, where we have set $u_1=4.5$, $u_2=0$ and $\Lambda=1$.}
	\label{table:asympweak1}
\end{table}

\begin{table}[h]
	\centering
	\begin{tabular}{|c|c|c|c|c|}
		\hline
		& \multicolumn{2}{|c|}{$\hbar=1$} & \multicolumn{2}{|c|}{$\hbar=0.7$} \\
		\hline 
		& evaluation (4.2-4.3) & $\frac{1}{\hbar}\Pi_{\gamma'}(\hbar)$ at $o(\hbar^2)$ & evaluation (4.2-4.3) & $\frac{1}{\hbar}\Pi_{\gamma'}(\hbar)$ at $o(\hbar^2)$\\
		\hline
		$\text{log}\cX_{\gamma'_1}$&  $13.411$ & $13.408$ & $19.1723$& $19.1715$ \\
		\hline
		$\text{log}\cX_{\gamma'_2}$&  $-13.411$ & $-13.408$ & $-19.1723$ & $-19.1715$\\
		\hline
		$\text{log}\cX_{\gamma'_3}$& $6.67+2.71\text{i}$ & $6.71+2.66\text{i}$& $9.61+3.89\text{i}$ & $9.59+3.88\text{i}$\\
		\hline
		$\text{log}\cX_{\gamma'_4}$ & $-6.67-2.71\text{i}$  & $-6.71-2.66\text{i}$& $-9.61-3.89\text{i}$& $-9.59-3.88\text{i}$\\
		\hline
	\end{tabular}
	\caption{Comparison of $\text{log}\cX_{\gamma'}$ with the formal quantum periods expansion up to order-$\hbar^2$ for equation $B$ of (\ref{eqn:SU3eqn}) at $\hbar=1$ and $\hbar=0.7$, where we have set $u_1=4.5$, $u_2=0$ and $\Lambda=1$.}
	\label{table:asympweak2}
\end{table}

In the evaluation of $\text{log}\cX_{\gamma'}$ using (\ref{eqn:Xweak1}) and (\ref{eqn:Xweak2}), similar to what happens in the strong coupling region, we found it difficult to perform numerical evaluation at small $\hbar$. Compared to the strong coupling region, here the numerical evaluation contains an extra step, namely diagonalizing the monodromy matrix $M$ and finding its eigenvalues and eigenvectors. As $\hbar$ gets rather small, one of the eigenvalues for $M$ becomes very small, introducing further difficulty in the numerical analysis. Nevertheless, from \autoref{table:asympweak1} and \autoref{table:asympweak2} one can see as $\hbar$ goes from $1$ to $0.7$, $\hbar\text{log}\cX_{\gamma'}$ becomes closer to the truncated quantum periods expansion. This provides evidence for the conjectured asymptotic behavior of spectral coordinates. 

\subsection*{Acknowledgement}

We thank Dylan Allegretti, Alba Grassi, Saebyeok Jeong, Dustin Lorshbough, Gregory Moore and Andrew Neitzke for very helpful discussions. FY is supported by DOE grant  DE-SC0010008.


\bibliographystyle{utphys}

\bibliography{SU3WKB}

\end{document}